\documentclass[1p]{elsarticle}

\usepackage{hyperref}
\usepackage{graphicx}
\usepackage{epstopdf}
\usepackage{multirow}
\usepackage{booktabs}
\usepackage{multicol, blindtext}
\usepackage{tabularx}
\usepackage{caption}
\usepackage{gensymb}
\usepackage{amsmath}
\usepackage{array}
\usepackage{subfig}
\usepackage{xcolor}
\usepackage{color, colortbl}
\usepackage{dblfloatfix}
\usepackage{babel} 
\usepackage{apalike}

\begin{document}

\begin{frontmatter}

\title{Classification of Motor faults Based on Transmission Coefficient and Reflection Coefficient of Omni-Directional Antenna Using DCNN}

\author[1]{Sagar Dutta}
\ead{sagar0dutta@gmail.com}
\author[1]{Banani Basu\corref{cor1}}%
\ead{banani@ece.nits.ac.in}
\author[1]{Fazal Ahmed Talukdar}
\ead{fazal@ece.nits.ac.in}
\cortext[cor1]{Corresponding author}

\address[1]{Department of Electronics and Communication Engineering \protect\\ National Institute Of Technology, Silchar, Assam, India}

\begin{abstract}
The most commonly used electrical rotary machines in the field are induction machines. In this paper, we propose an antenna based approach for the classification of motor faults in induction motors using the reflection coefficient $S_{11}$ and the transmission coefficient $S_{21}$ of the antenna. The spectrograms of $S_{11}$ and $S_{21}$ is seen to possess unique signatures for various fault conditions that are used for the classification. To learn the required characteristics and classification boundaries, deep convolution neural network (DCNN) is applied to the spectrogram of the S-parameter. DCNN has been found to reach classification accuracy 93\% using $S_{11}$, 98.1\% using $S_{21}$ and 100\% using both $S_{11}$ and $S_{21}$. The effect of antenna operating frequency, its location and duration of signal on the classification accuracy is also presented and discussed.
\end{abstract}

\end{frontmatter}


\section{Introduction}

Mechanical machinery is getting more advanced and intelligent with the rapid growth of science and technology and the ongoing development of industrial applications. Rotating parts, which are critical components of machines, are commonly used in industries, and their failure may lead to severe losses and disastrous effects. As per the EPRI (Electric Power Research Institute) reports \citep{one}, 41\% of the induction motor failure is caused by bearing defects, 9\% by rotor failures and 36\% by stator failures. A condition monitoring and fault diagnosis system \citep{z1,z2,z3} is necessary as it is easier to maintain the induction motor and replace the faults, rather than allowing the faulty motor to shut down the operation. For safe and efficient operation of the rotating machinery, proper maintenance and condition monitoring is necessary. To address this issue, various parameters of the motor such as vibration, temperature, sound, motor current, acoustic emission and stray flux around the motor are monitored to identify faults.

\begin{table*}[!t]
\centering
\caption{Different fault analysis methods}
\resizebox{\textwidth}{!}{%

\begin{tabular}{@{}llll@{}}
\toprule
Method                         & Faults identified                                                                           & Measured Signal                 & Sensor used                                                                                              \\ \midrule
Eddy Current Analysis \citep{two}          & Inner Race                                                                                  & Vibration                       & Eddy Current Sensor                                                                                      \\ \\
Sound Analysis \citep{eight}        & \begin{tabular}[c]{@{}l@{}}Broken Rotor Bar,\\ Bearing Defect \\ and Unbalance\end{tabular} & Sound                           & Condensor Microphone                                                                                     \\ \\
Motor Current Signature \citep{nine}        & Bearing Defect                                                                              & Current                         & Current Sensor                                                                                           \\ \\
Acoustic emission \citep{six}              & split-torque gearbox, gear seeded fault                                                     & Acoustic emission signal        & AE sensors                                                                                               \\ \\
Flux Analysis \citep{ten}         & Bearing Defect                                                                              & Stray Flux                      & Magnetic Flux Sensor                                                                                     \\ \\
Optical Doppler Shift \citep{fourteen} & Test Speaker Vibrations                                                                     & Vibration                       & Laser Module                                                                                             \\ \\
Phase Locked Loop \citep{fifteen}        & Rotor and Bearing Defect                                                                    & Vibration                       & Ultra Wideband Radar                                                                                     \\ \\
Multi Sensor Wireless \citep{vca}          & Bearing Defect and AirGap Eccentricity                                                      & Acoustic, Vibration and Current & \begin{tabular}[c]{@{}l@{}}Hall Effect Sensor, \\ Accelerometer (2-axial) \\ and Microphone\end{tabular} 
\\ \\
Speed-based \citep{rotor}          & Bearing Defect                                                      & Rotor speed & \begin{tabular}[c]{@{}l@{}}E60H NPN Type Rotary Encode\end{tabular} \\ \bottomrule

\end{tabular}}
\label{comparison}
\end{table*}

Xue et. al \citep{two} proposed a motion induced eddy current sensor that can measure vibration signal in a non-intrusive manner of a non-ferromagnetic object. The bearing cage temperature and vibration signal are simultaneously measured at different operating speeds \citep{inductive}, the temperature is sensed by measuring the thermal-induced shift in the resonant frequency of an inductive coil with the help of another coil placed near it. Spectral analysis \citep{eight} is applied to the sound produced by the induction motor to identify different faults based on MUSIC algorithm. The motor current signal analysis is a popular method for diagnosing faults in induction motors \citep{nine} which is non-invasive and easy to implement. The acoustic emission of the induction motor is analyzed \citep{six} to monitor machine health and diagnose faults. They are typically unaffected by the background mechanical noise and are sensitive to the location of the faults. The stray flux radiated by the motor around it can be affected by the faults in the machine, which can be used for fault diagnosis \citep{ten}. Laser Doppler Vibrometry (LDV) is another technique that is used for non-contact vibration measurement. It is based on the principle of doppler frequency shift of the reflected light by a vibrating object \citep{fourteen}. A handheld doppler ultra wide band RADAR is used in \citep{fifteen} to diagnose multiple bearing faults in a squirrel cage induction motor non invasively. A combination of sensors such as acoustic, current and vibration sensor is utilized to detect fault using wireless network \citep{vca}. Rotor speed-based method \citep{rotor} is utilized to diagnose bearing faults.

The vibration, current, and acoustic emissions are measured using a network of wireless sensors to detect internal and external bearing race defects. Different techniques of wireless sensor network is used to identify bearing faults by measuring the motor parameters such as accelerometer \citep{acce1,acce2}, induction coupling \citep{inductive}, and amplitude modulation \citep{am}. However, these methods require complex analysis, high processing time and temperature correction circuit to avoid overheating of the sensors. Most of the sensors used are intrusive and thus requires the disruption of operation for their installation and maintenance which causes loss financially.

Furthermore, the interplay between optimization and machine learning is one of the most important developments in modern computational science. Optimization formulations and methods are proving to be vital in designing algorithms to extract essential features. Optimizer such as Bat algorithm \citep{g1}, Harris Hawk \citep{g2}, Whale optimizer \citep{g3}, Vortex search algorithm \citep{g4}, Spotted Hyena \citep{g5} and Symbiotic Organisms Search algorithms \citep{g6} can be implemented in machine learning to improve it's performance.

The objective of this paper is to present a new approach of measuring vibration signals using reflection coefficient ($S_{11}$) and transmission coefficient ($S_{21}$) \citep{balani} of an omni-directional antenna. This is a cost-effective and robust approach that does not require any conventional sensor and is non-invasive. As the impedance of an antenna and the transmission path between two antennas can be disrupted by any vibrating object, therefore they can be used to identify different vibration signals due to specific faults. We introduce a deep convolutional neural network to classify bearing faults (inner and outer race) and rotor imbalance based on the spectrogram of the measured S-parameters. In this proposed method, the antennas are placed near the vibration site, and the reflection coefficient and the transmission coefficients are measured. The vibration induced by the fault causes a variation in the $S_{11}$ and $S_{21}$. The performance of classification based on the S-parameters using DCNN is investigated. We have also explored the use of S-parameter in the classification of human activities \citep{sagar}.

According to the literature review, previous researchers have not used the antenna's near field effect and transmission pathloss to detect faults from vibrations. This study is one of the first to use an antenna as a sensor to detect vibrations caused by bearing faults and imbalance in induction motors. It is observed that the antenna’s reactive near field and transmission pathloss exhibits unique time-varying signatures and can thus potentially be used for fault analysis. Fast Fourier transform has been further used to extract the time-frequency domain features. Moreover, we implemented a deep learning model that effectively classifies different faults based on the aforesaid method to demonstrate the capability of our approach to integrate with the ongoing fourth industrial revolution and smart technology, and the results demonstrate its potential.

The paper is organized as follows. Section II presents the experimental setup and measurement. The classification technique is introduced in Section III. Results are presented and discussed in Section IV. Section V summarizes the paper.

\begin{figure}[!h]
    \centering
    \includegraphics[width=3.5 in]{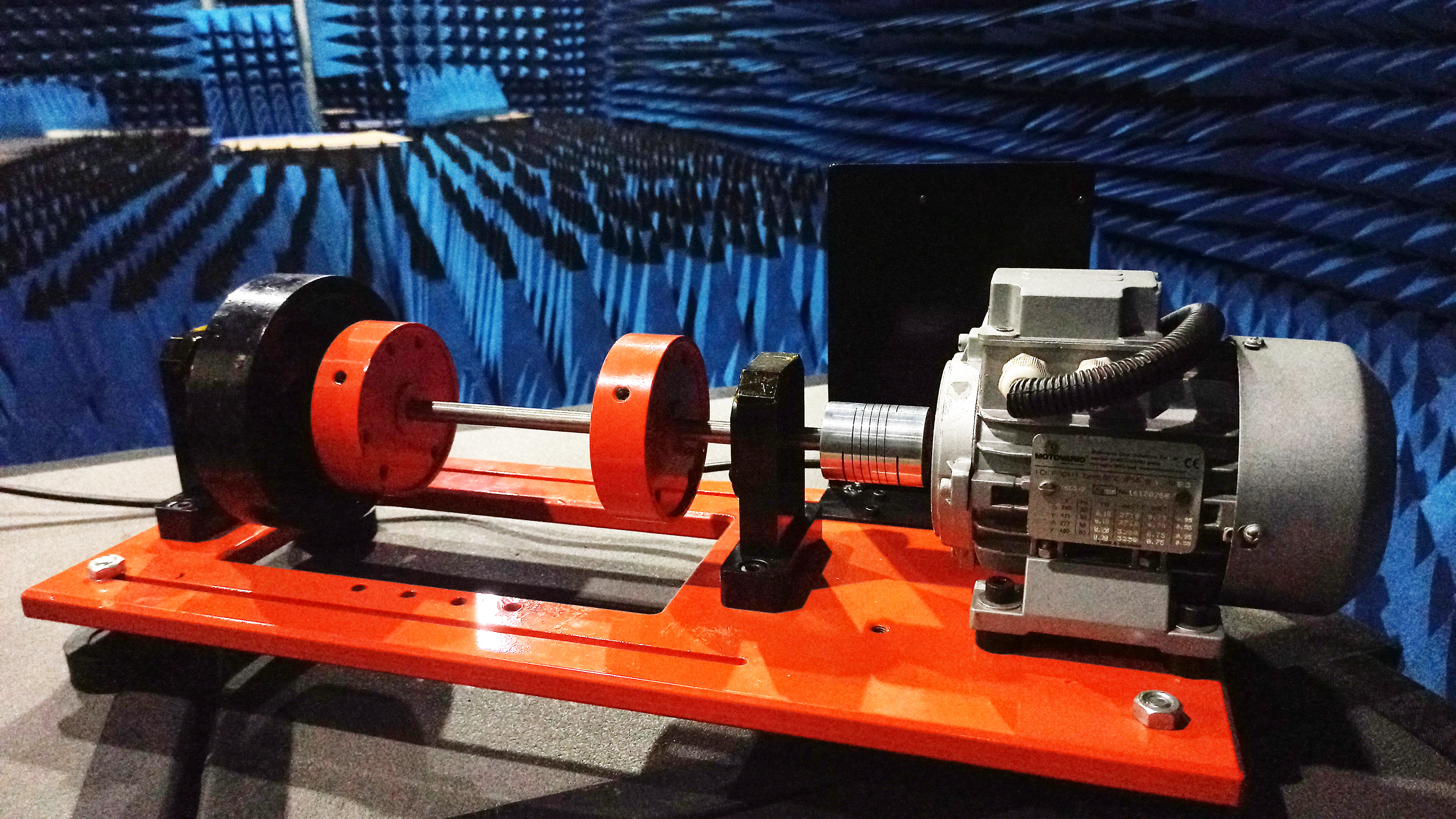}
    \caption{Experimental Setup}
    \label{setup}
\end{figure}

\section{Measurement of S-parameters}
The experimental setup used in our study is shown in Fig. \ref{setup}. It comprises a 3-phase AC induction motor, a shaft of length 450 mm, and a diameter of 12 mm, which is coupled with a helical coupler. The shaft is supported by two bearing blocks that house the ball bearings. A load disk (black) is used as a static load, and a pair of balance disk (orange) measuring 30 mm in thickness and 200 mm in diameter is used to introduce static imbalance in the setup.

\begin{figure}[!t]
\centering
\subfloat[]{\includegraphics[width=1.5 in]{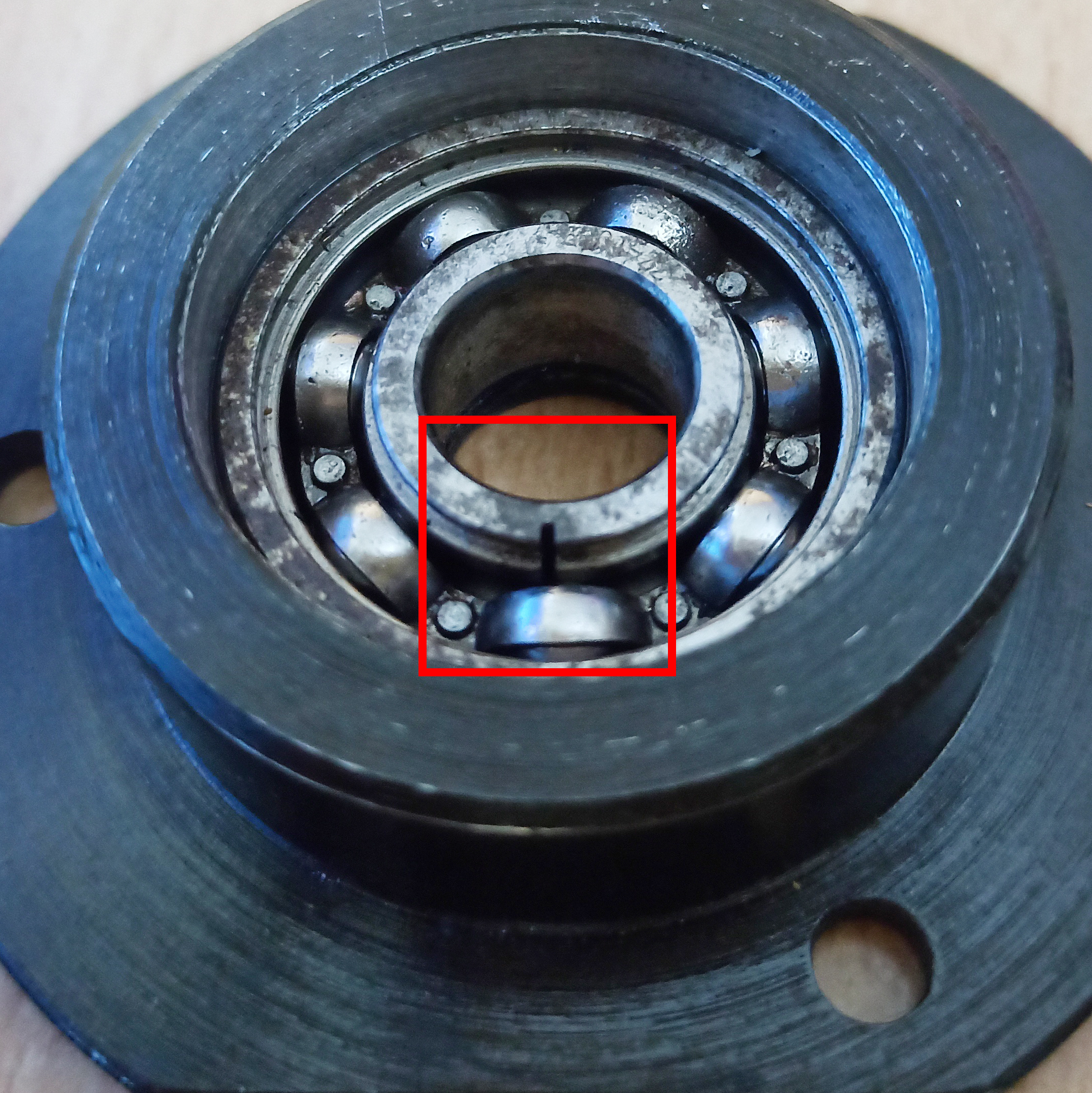}%
\label{bearing_1a}}
\hfill
\subfloat[]{\includegraphics[width=1.5 in]{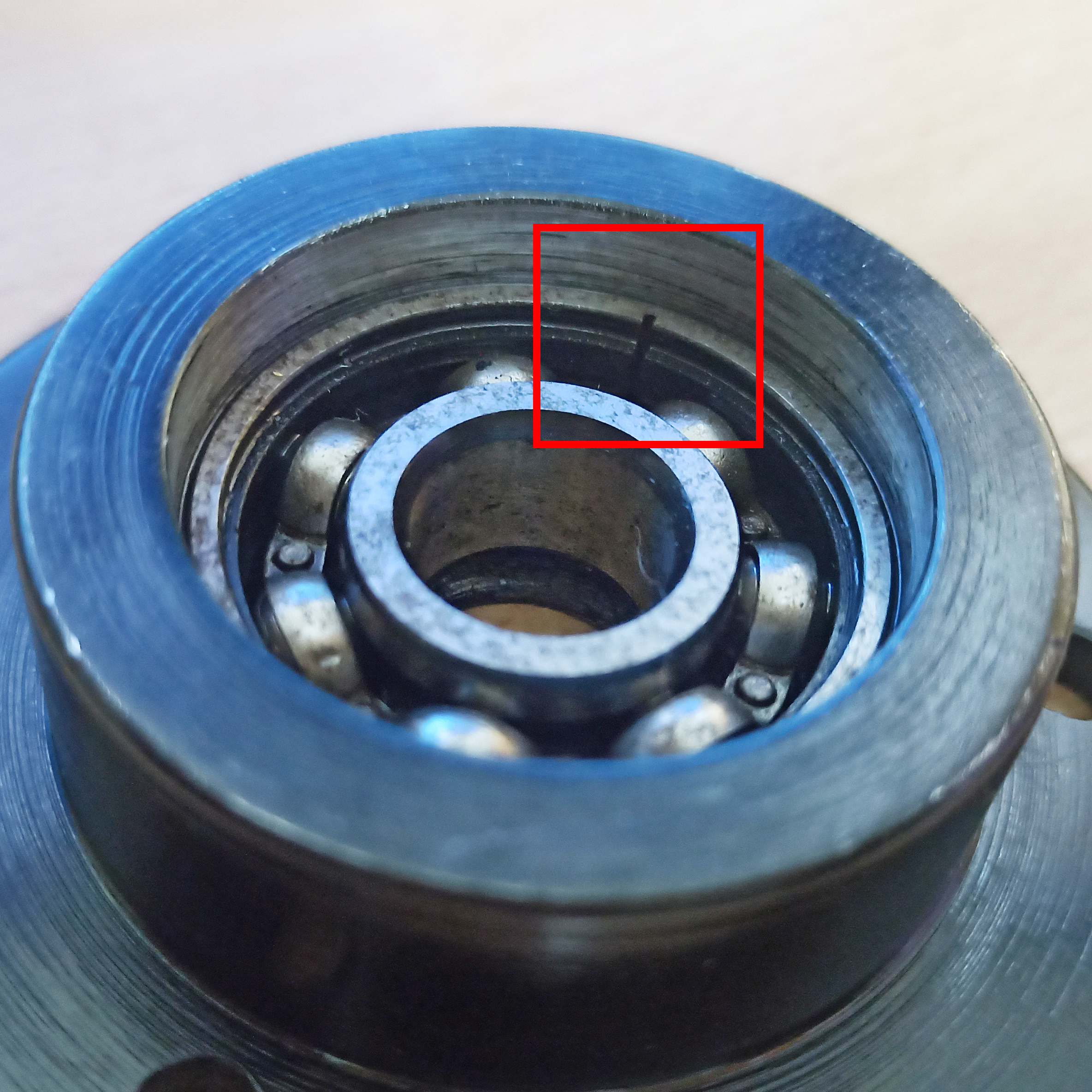}%
\label{bearing_1b}}
\hfill
\subfloat[]{\includegraphics[width=1.5 in]{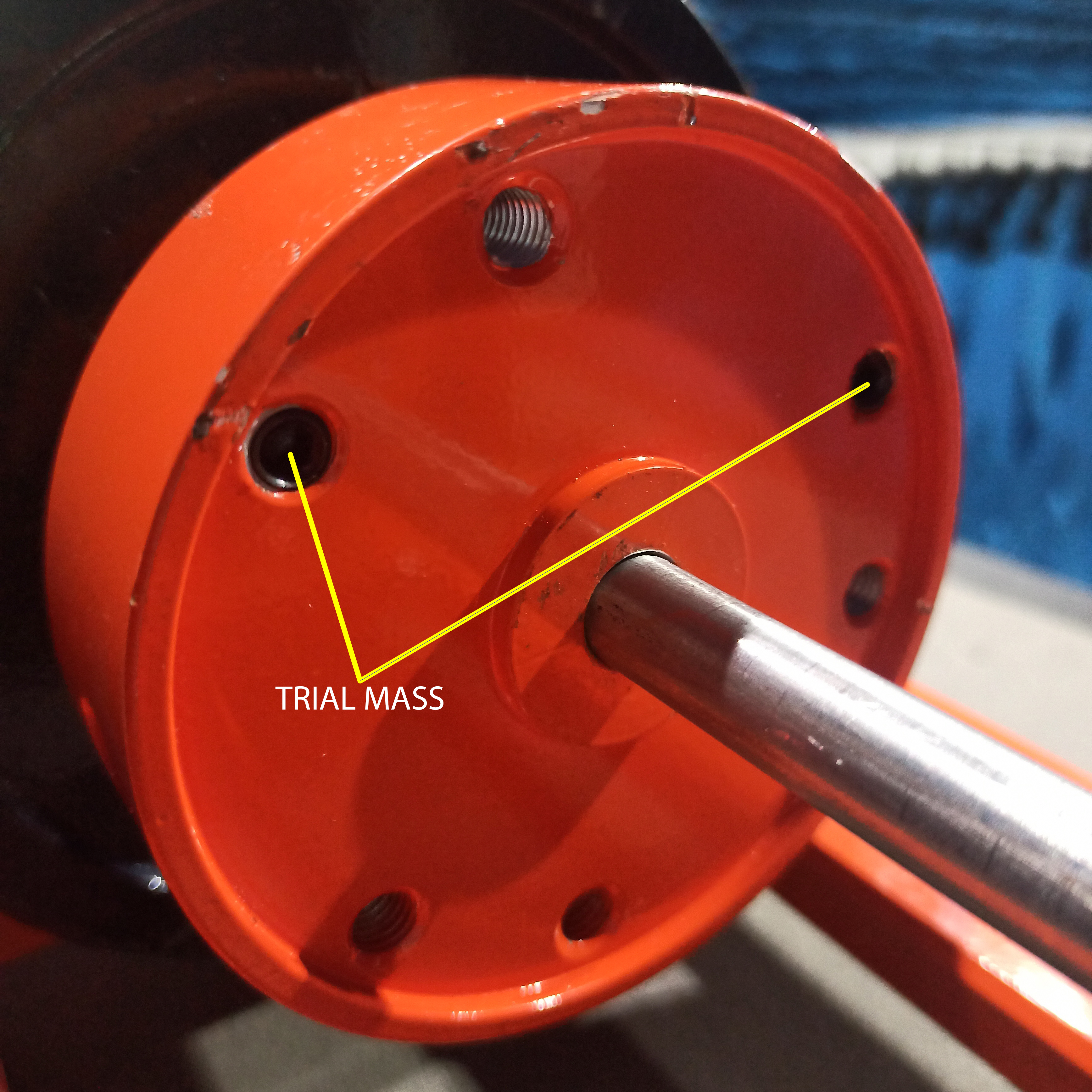}%
\label{bearing_1c}}
\caption{(a) Inner race fault  (b) Outer race fault (b) Imbalance disk with trial mass}
\label{bearing}
\end{figure}

The defects are artificially created in the ball bearings to analyze the vibration signal corresponding to the faults. The inner race and outer race defects in the ball bearings are created as shown in Fig. \ref{bearing_1a} and Fig. \ref{bearing_1b}. The imbalance in the setup is introduced by attaching trial masses to the disk as shown in Fig. \ref{bearing_1c}. The imbalance is caused by the shift in the center of gravity when the trial mass is added.

\begin{figure}[!t]
\centering
\subfloat[]{\includegraphics[width=2.3in]{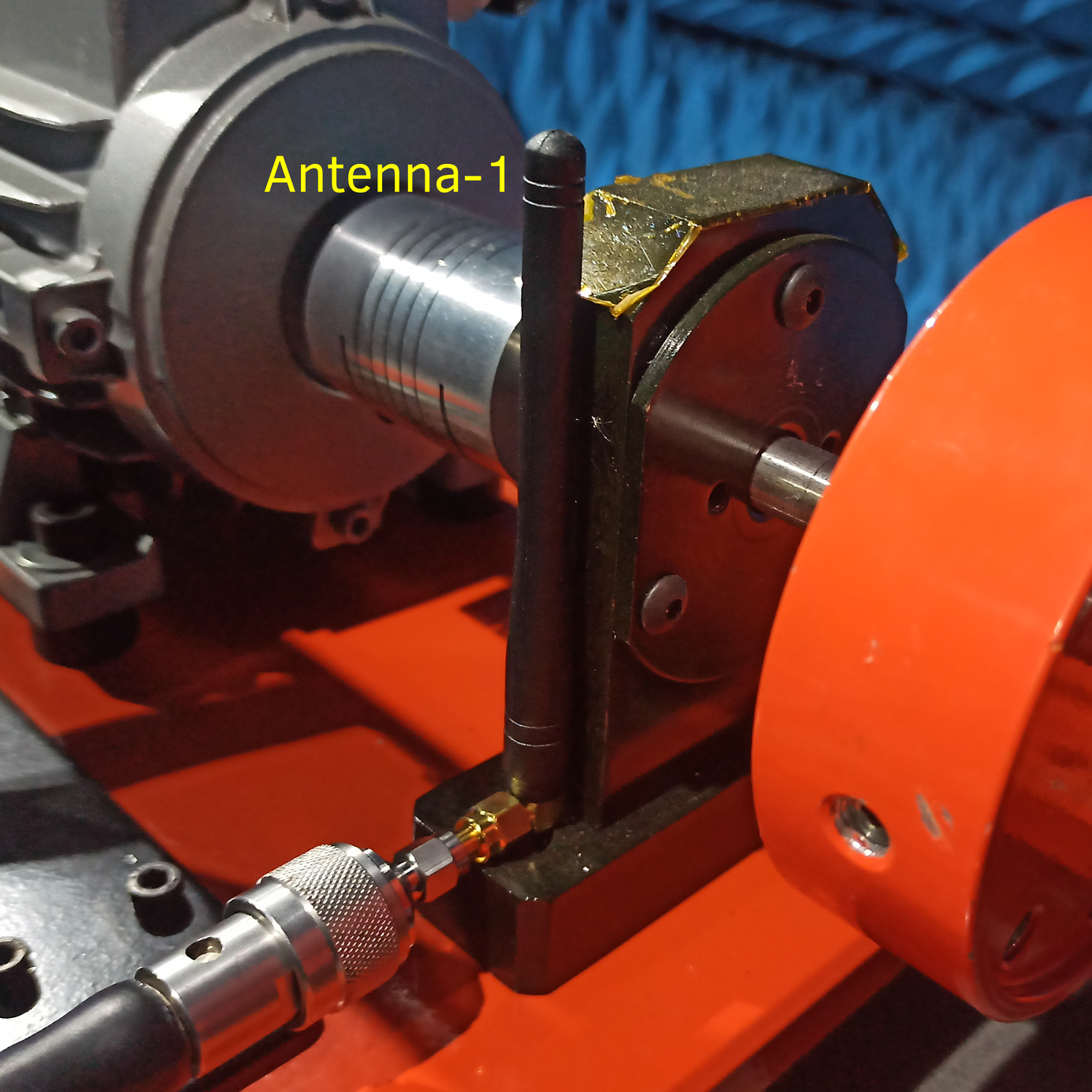}%
\label{setup_1a}}
\hfill
\subfloat[]{\includegraphics[width=2.3in]{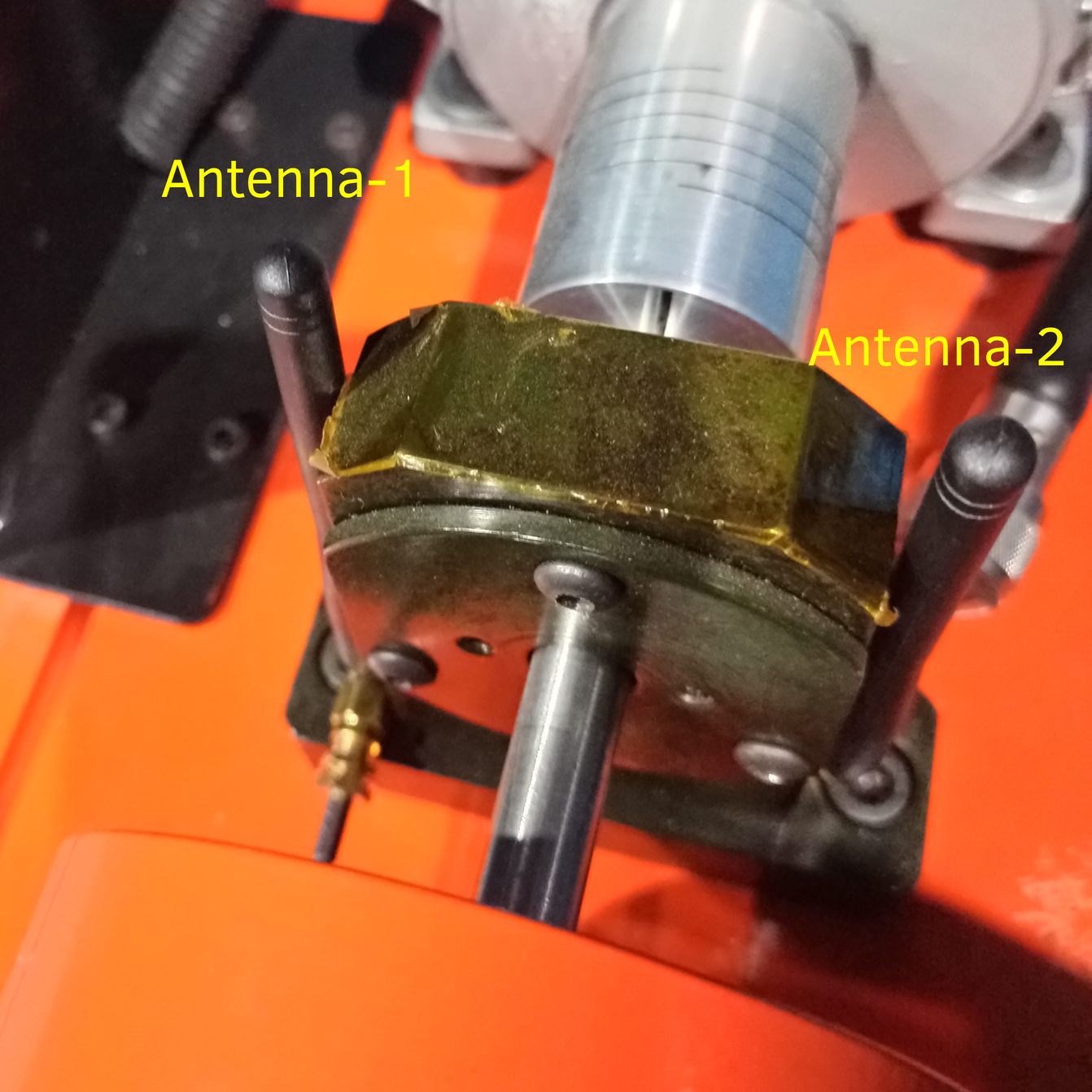}%
\label{setup_1b}}
\caption{(a) Placement of antenna for reflection coefficient measurement (b) Placement of antenna for transmission coefficient measurement}
\label{setup1}
\end{figure}

\begin{figure}[!h]
\centering
\subfloat[]{\includegraphics[width=1.5in]{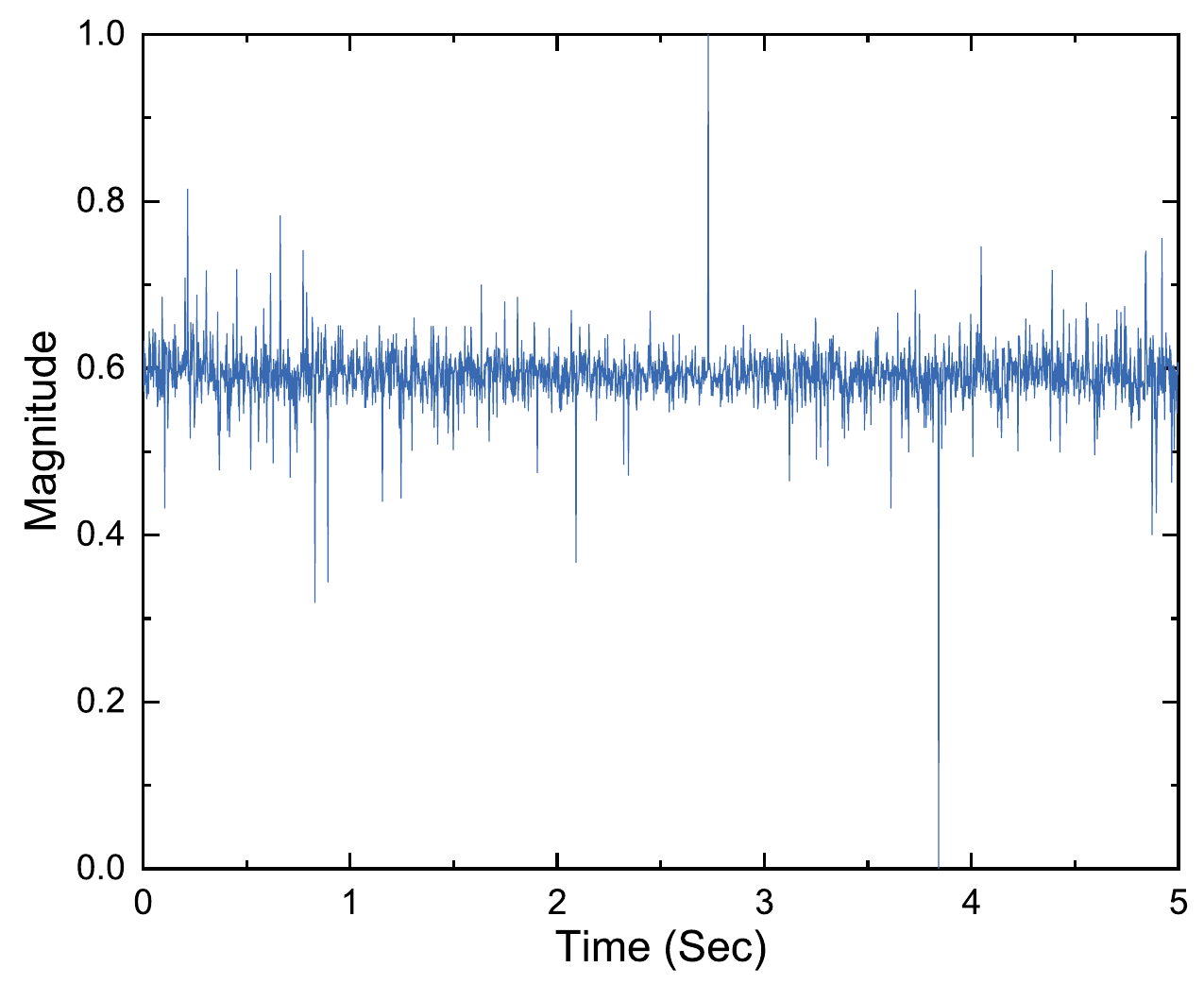}%
\label{raw11_1a}}
\hspace{5mm}
\subfloat[]{\includegraphics[width=1.5in]{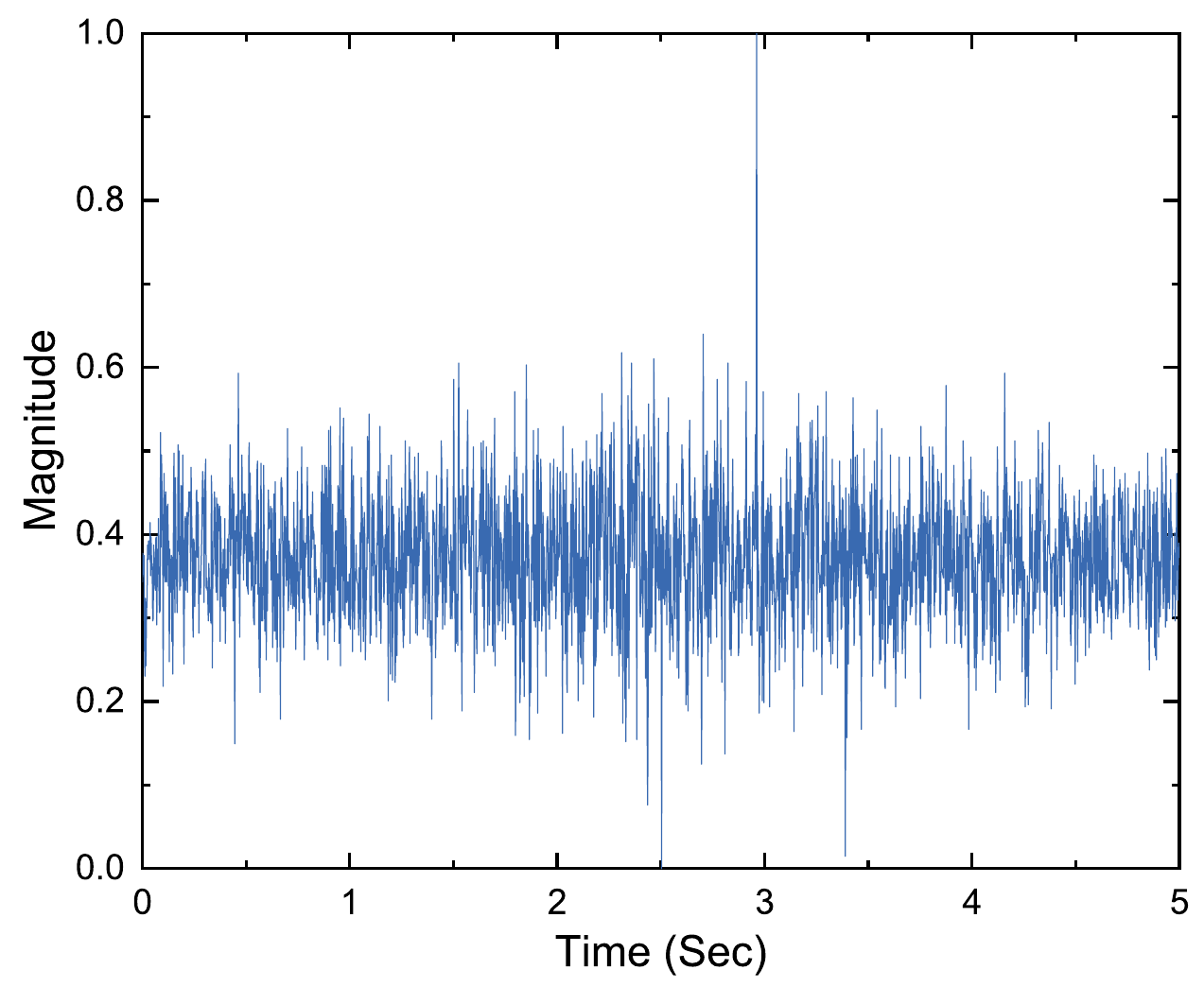}%
\label{raw11_1b}}

\subfloat[]{\includegraphics[width=1.5in]{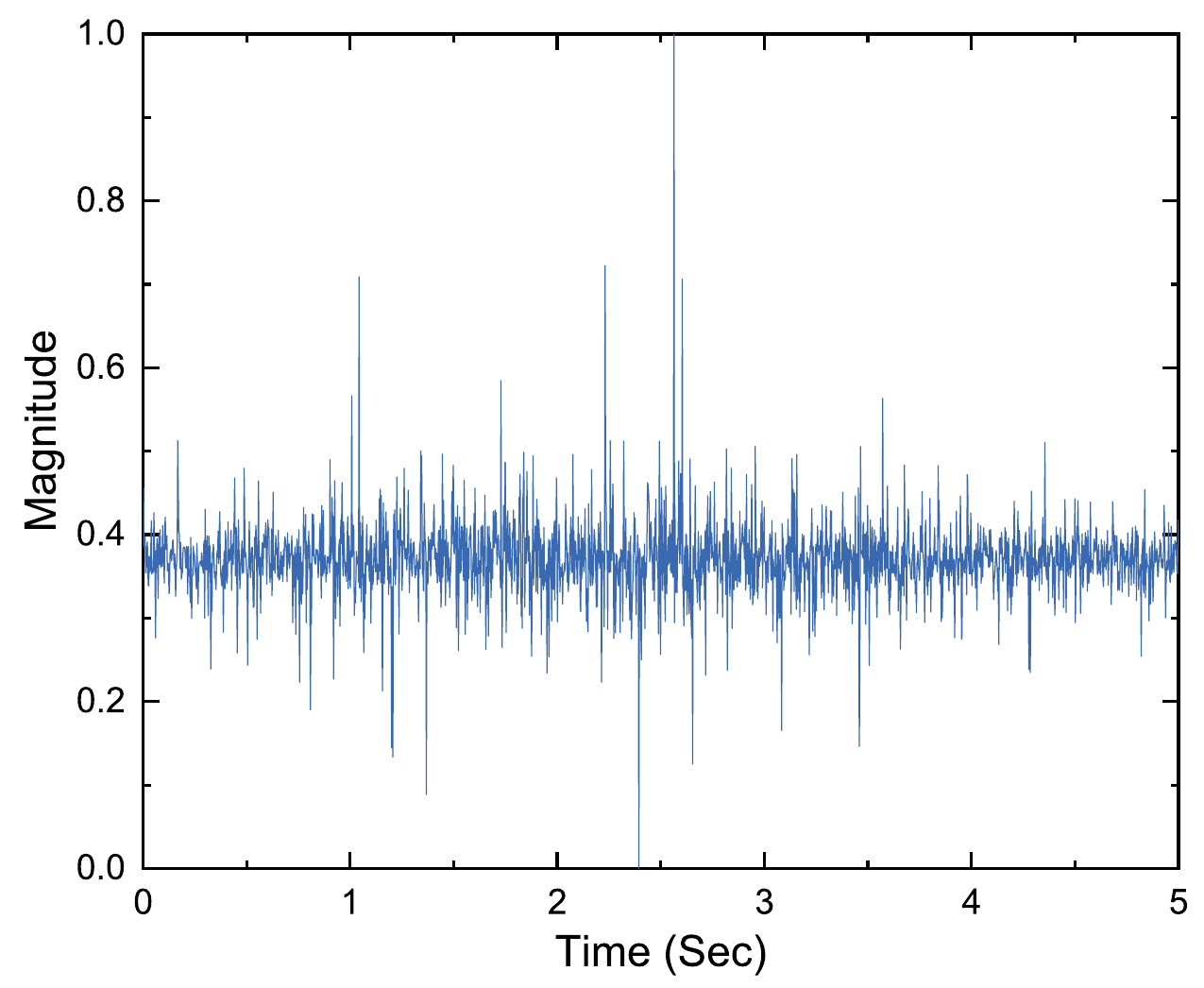}%
\label{raw11_1c}}
\hspace{5mm}
\subfloat[]{\includegraphics[width=1.5in]{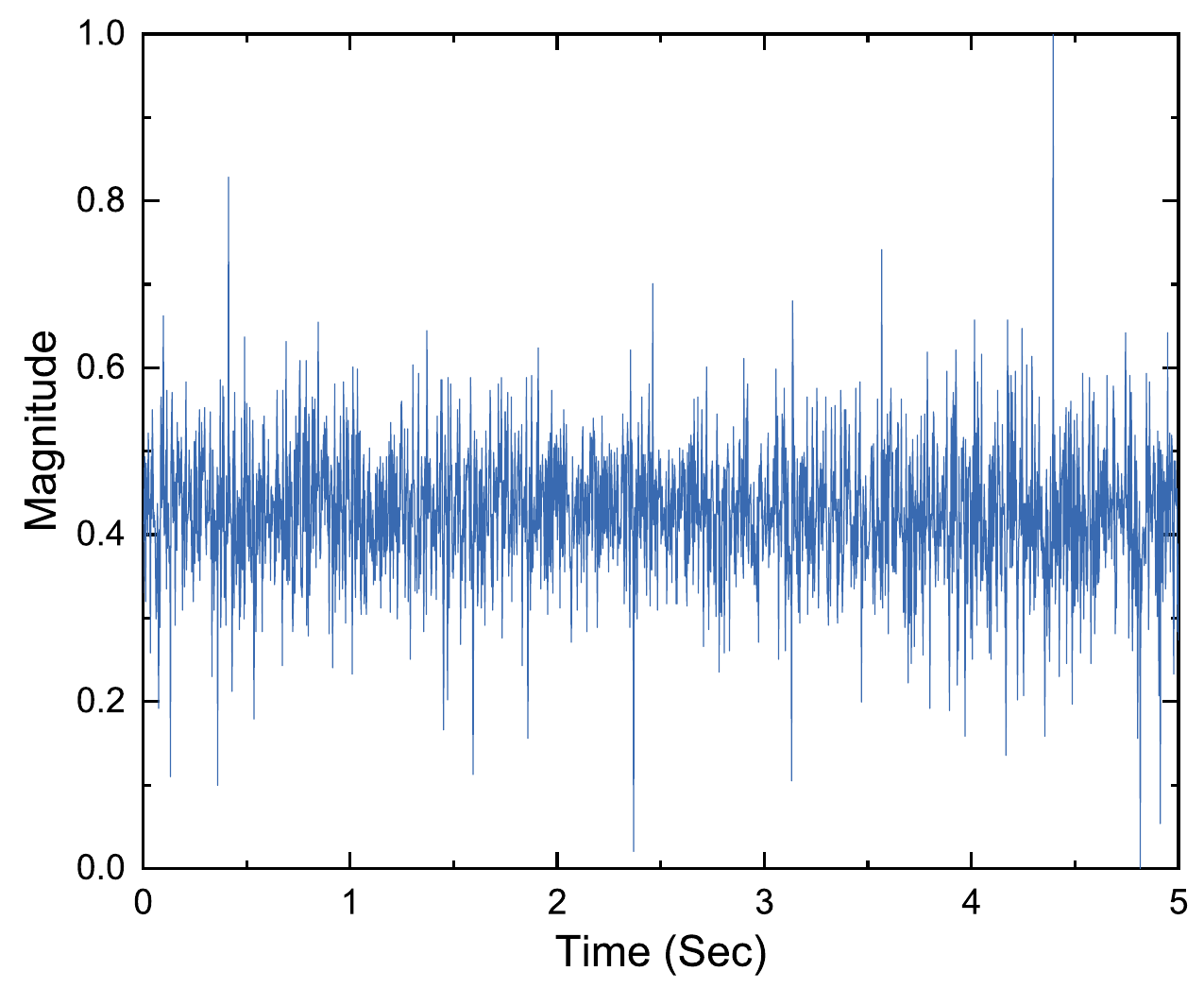}%
\label{raw11_1d}}
\caption{$S_{11}$ for different fault conditions: (a) Normal  (b) Imbalance (c) Inner race fault (d) Outer race fault}
\label{raw11}
\end{figure}

\begin{figure}[!h]
\centering
\subfloat[]{\includegraphics[width=1.5in]{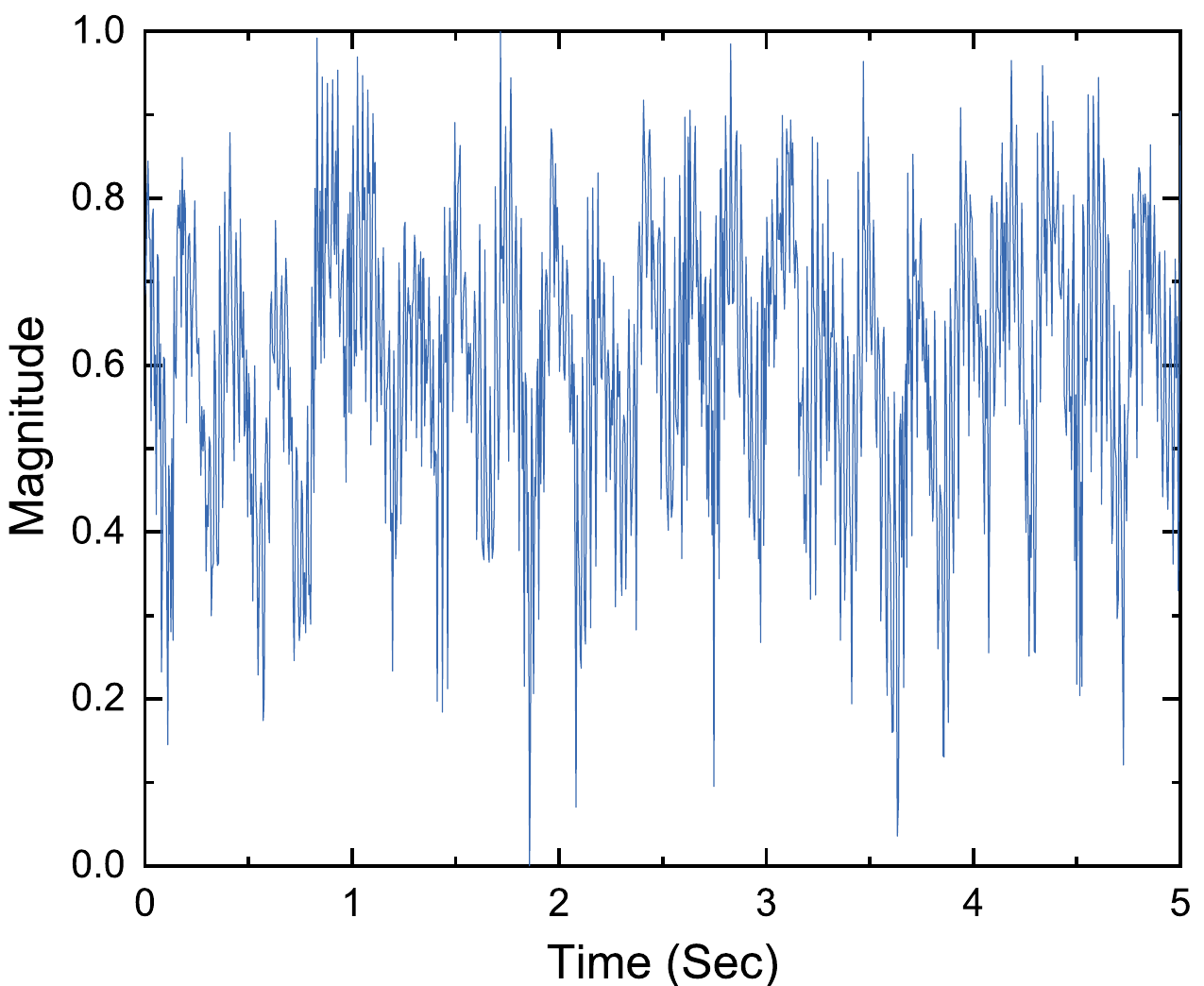}%
\label{raw21_1a}}
\hspace{5mm}
\subfloat[]{\includegraphics[width=1.5in]{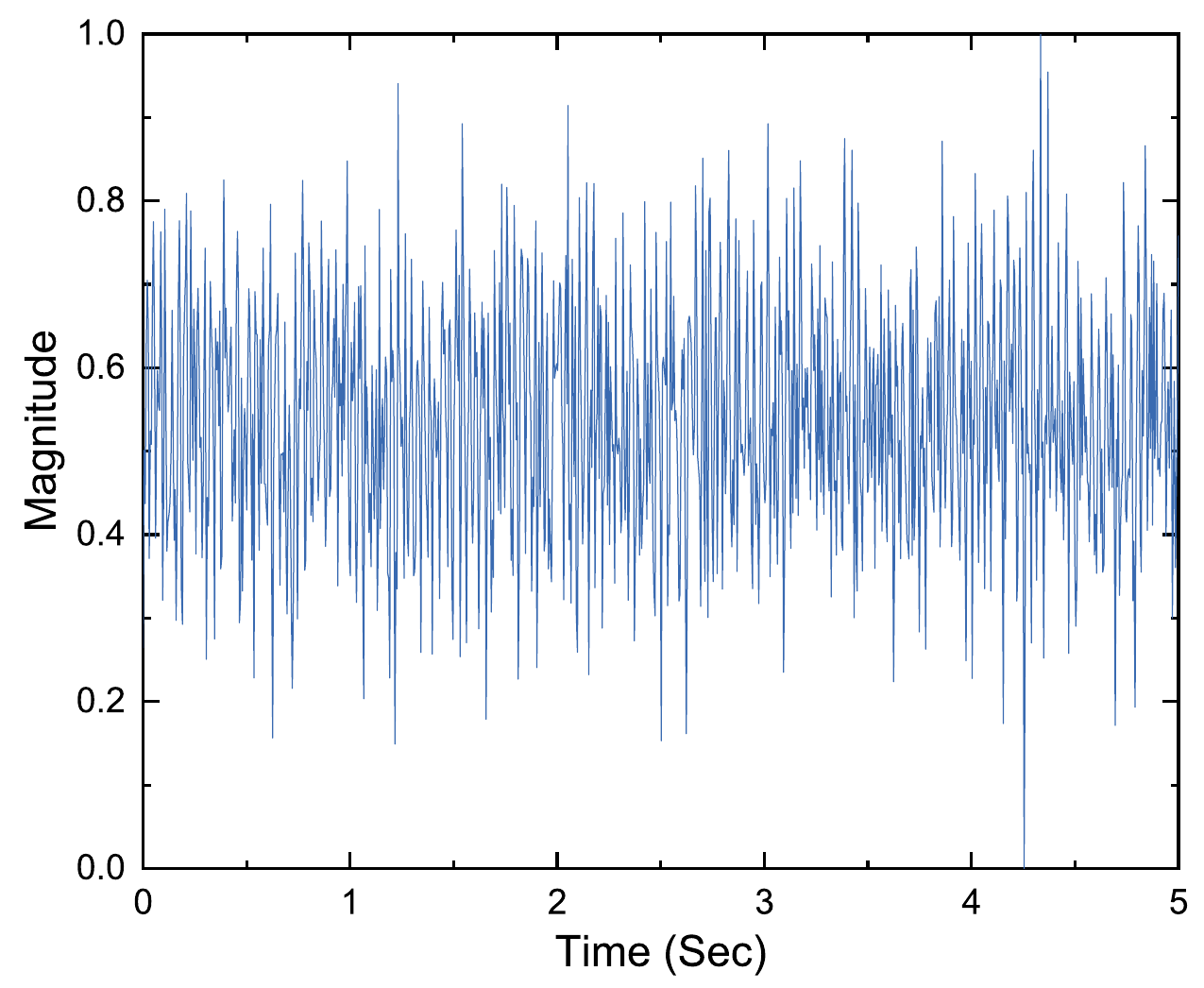}%
\label{raw21_1b}}

\subfloat[]{\includegraphics[width=1.5in]{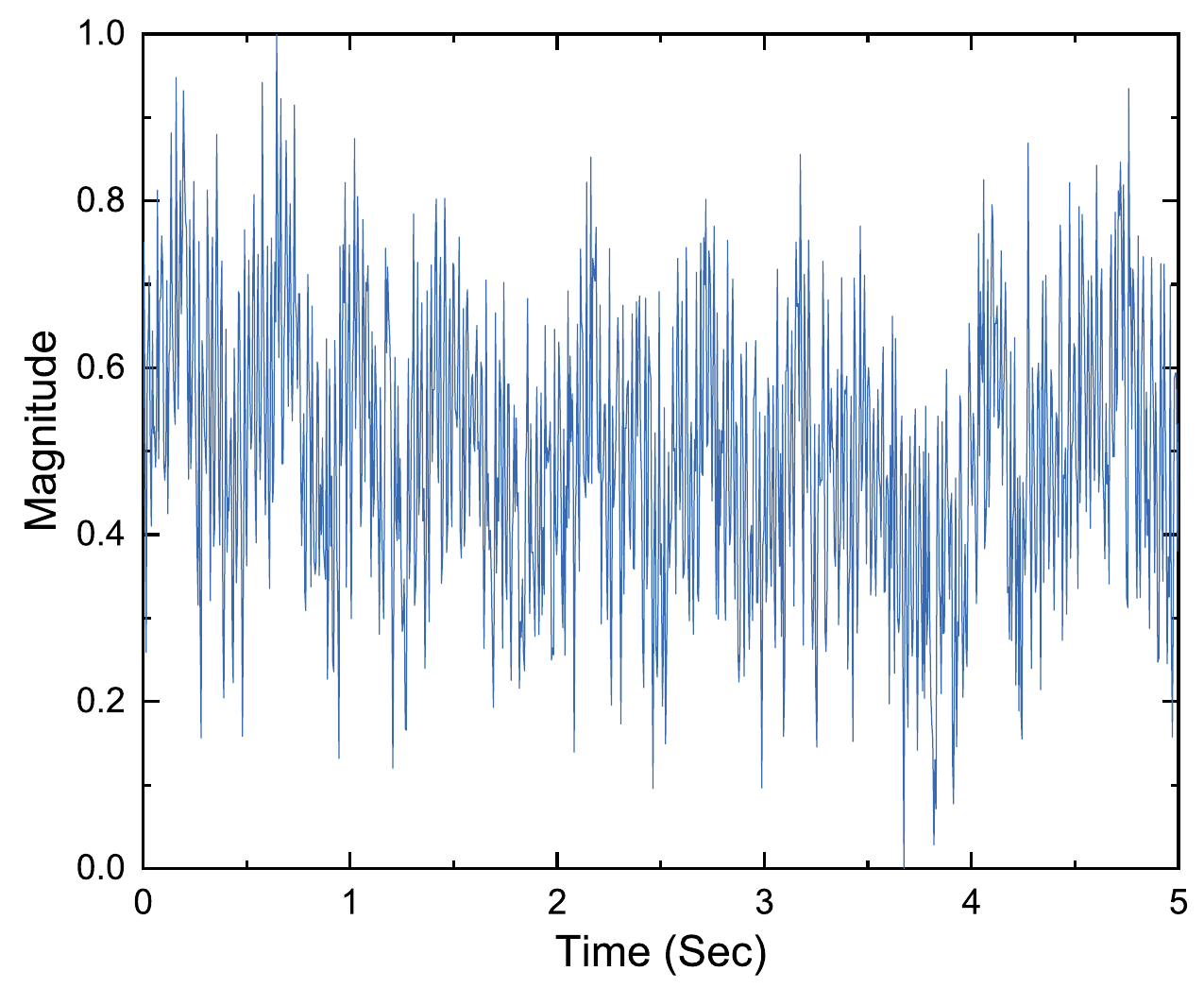}%
\label{raw21_1c}}
\hspace{5mm}
\subfloat[]{\includegraphics[width=1.5in]{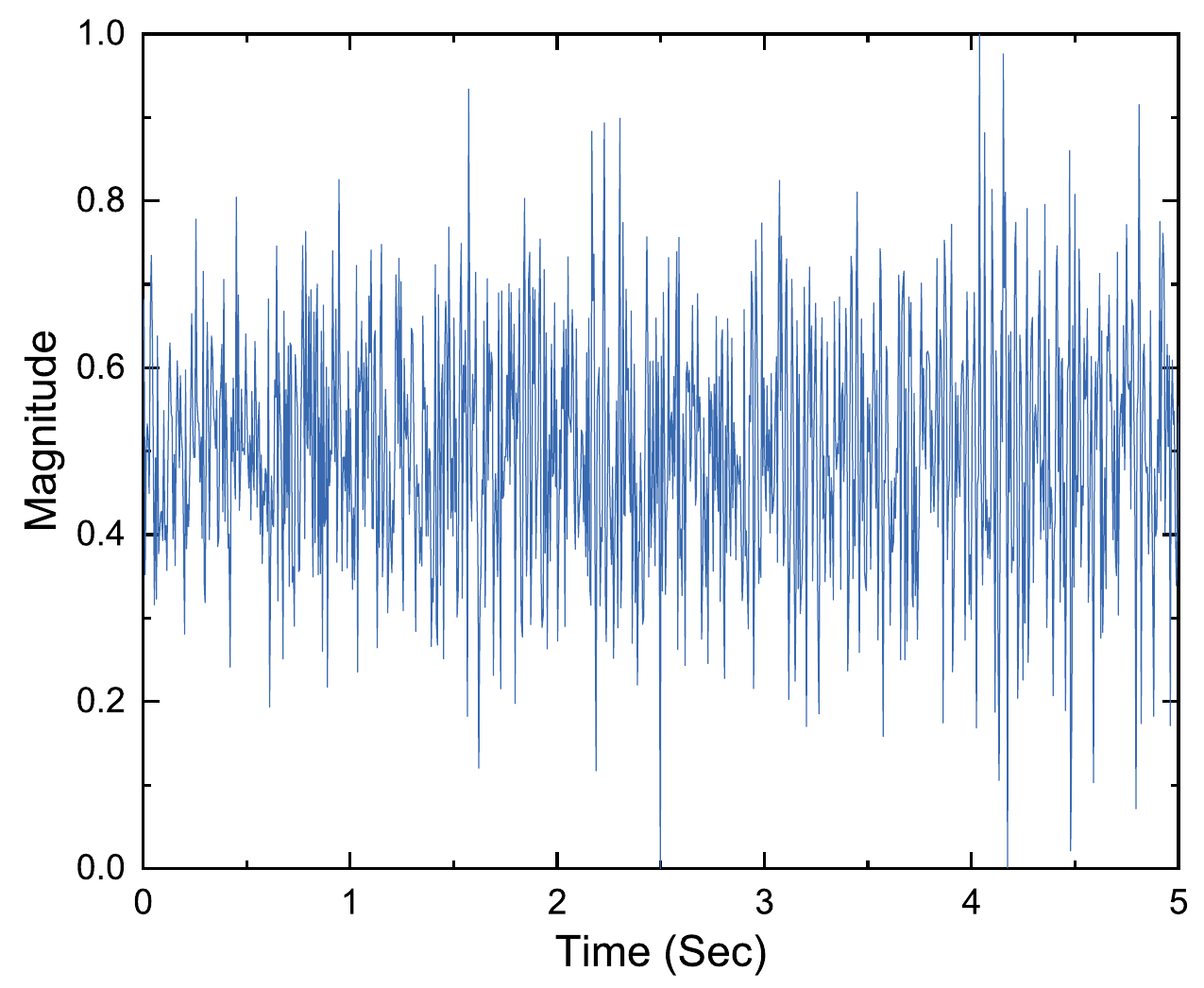}%
\label{raw21_1d}}
\caption{$S_{21}$ for different fault conditions: (a) Normal  (b) Imbalance (c) Inner race fault (d) Outer race fault}
\label{raw21}
\end{figure}


\begin{figure}[!h]
\centering
\subfloat[]{\includegraphics[width=1.6in]{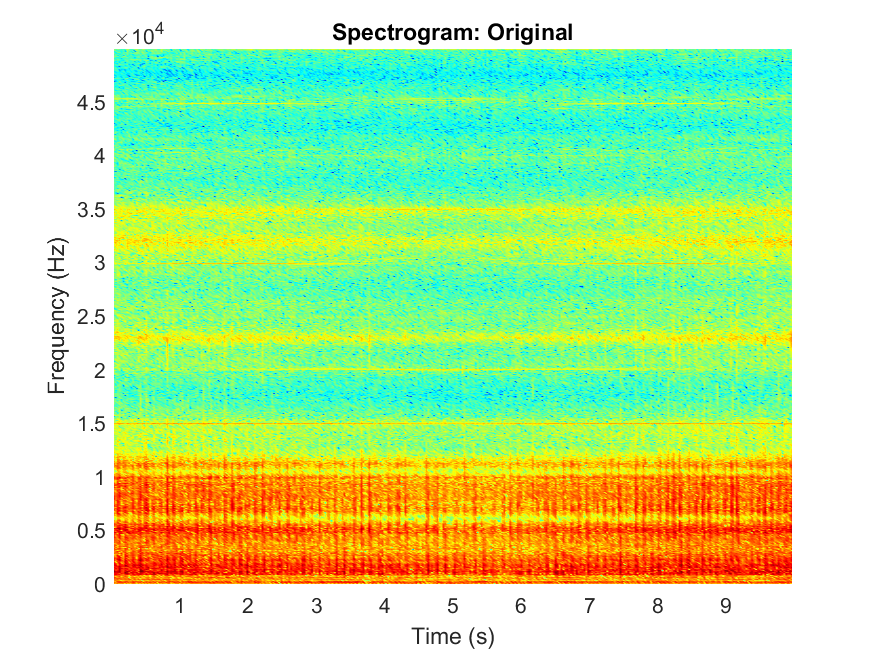}%
\label{spec11_1a}}
\hspace{5mm}
\subfloat[]{\includegraphics[width=1.6in]{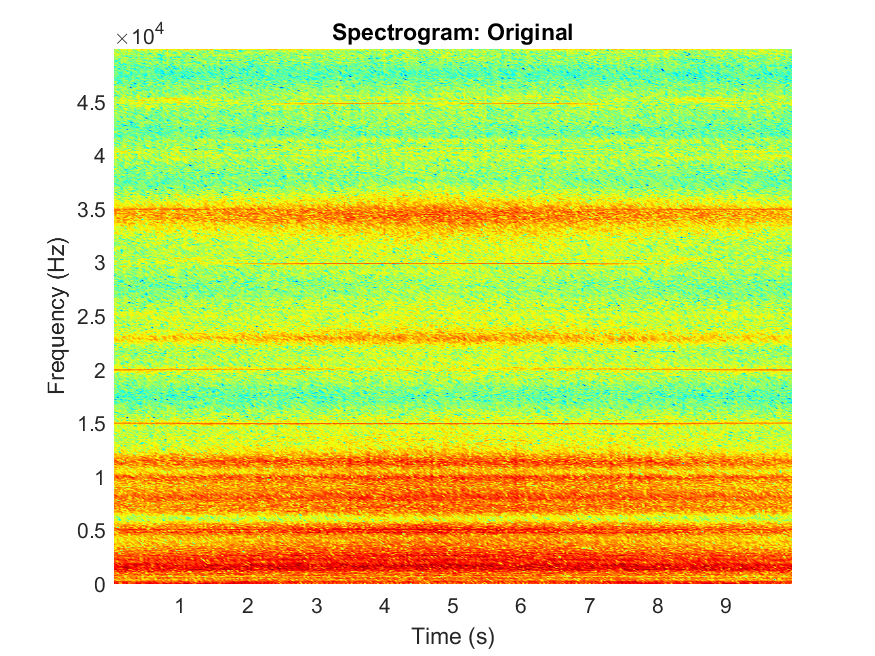}%
\label{spec11_1b}}

\subfloat[]{\includegraphics[width=1.6in]{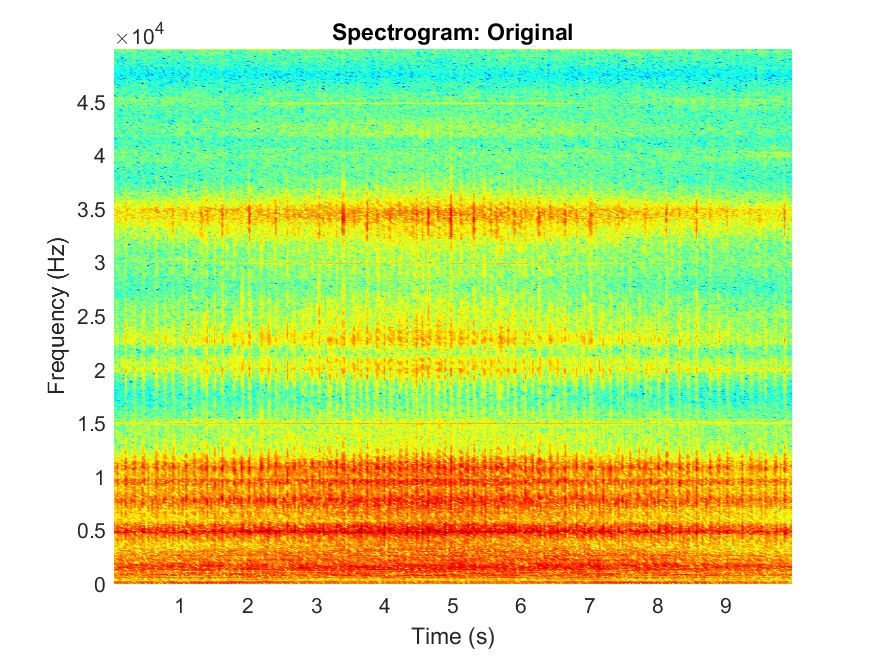}%
\label{spec11_1c}}
\hspace{5mm}
\subfloat[]{\includegraphics[width=1.6in]{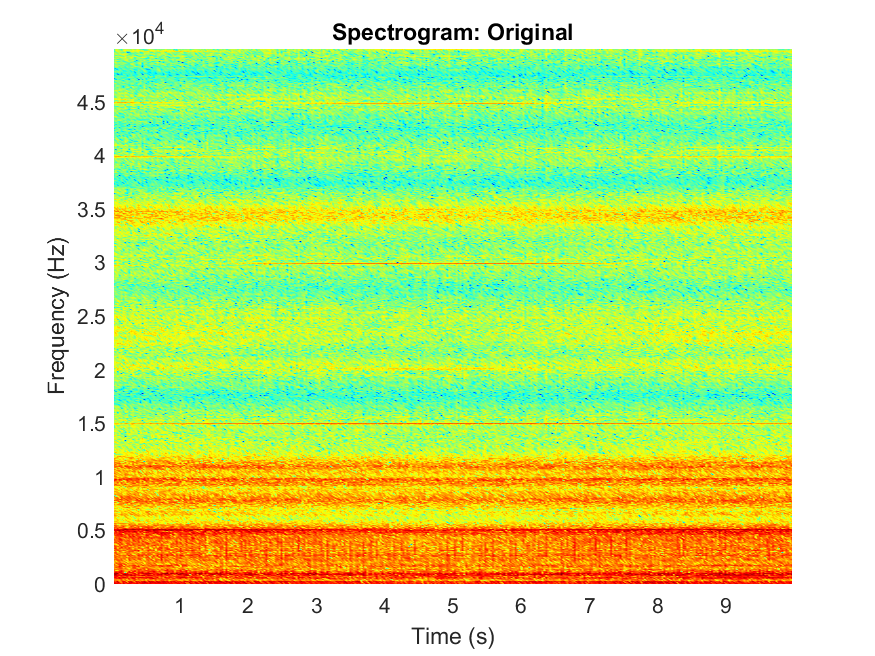}%
\label{spec11_1d}}
\caption{Spectrogram of $S_{11}$ for different fault conditions: (a) Normal  (b) Imbalance (c) Inner race fault (d) Outer race fault}
\label{spec11}
\end{figure}


\begin{figure}[!h]
\centering
\subfloat[]{\includegraphics[width=1.6in]{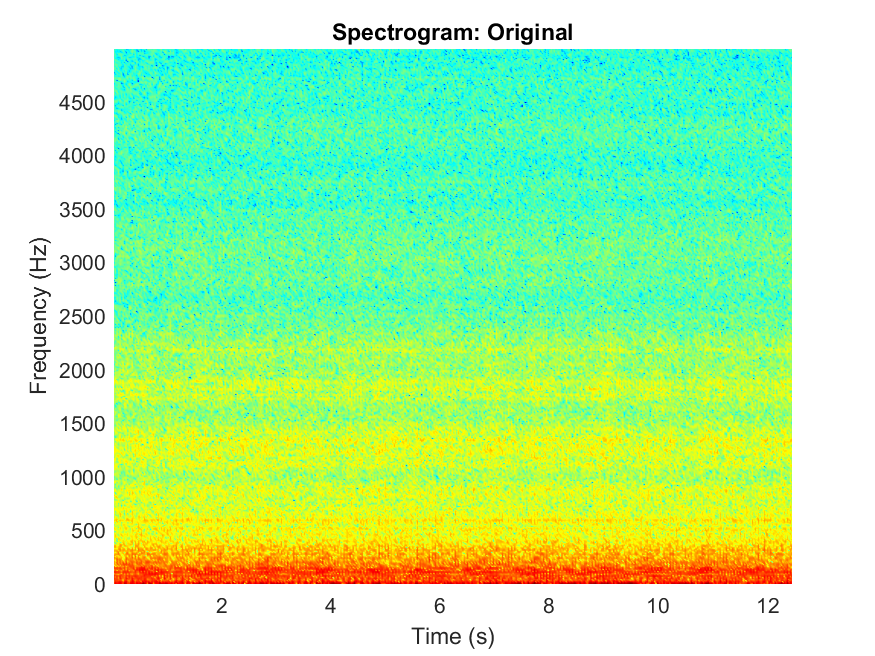}%
\label{spec21_1a}}
\hspace{5mm}
\subfloat[]{\includegraphics[width=1.6in]{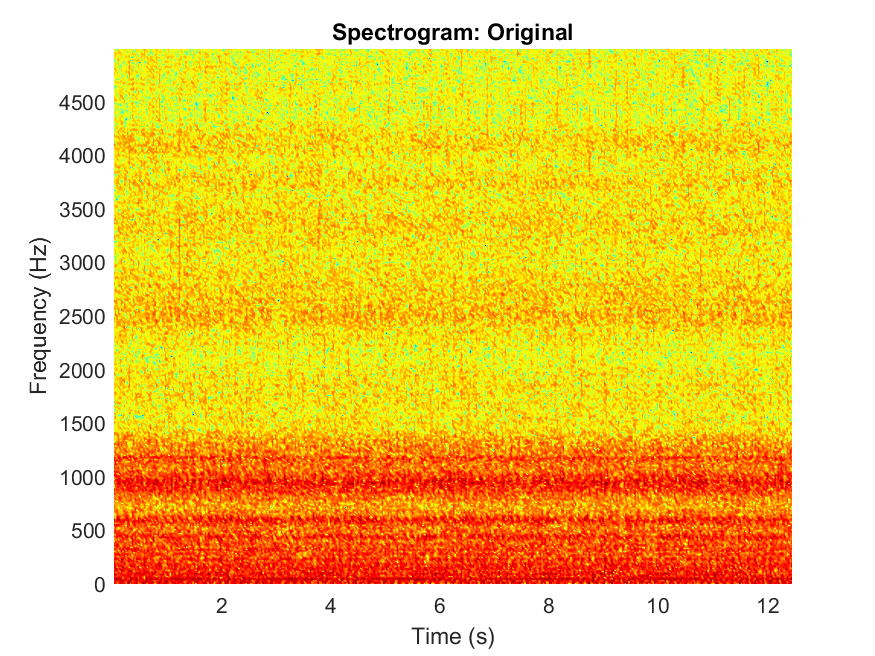}%
\label{spec21_1b}}

\subfloat[]{\includegraphics[width=1.6in]{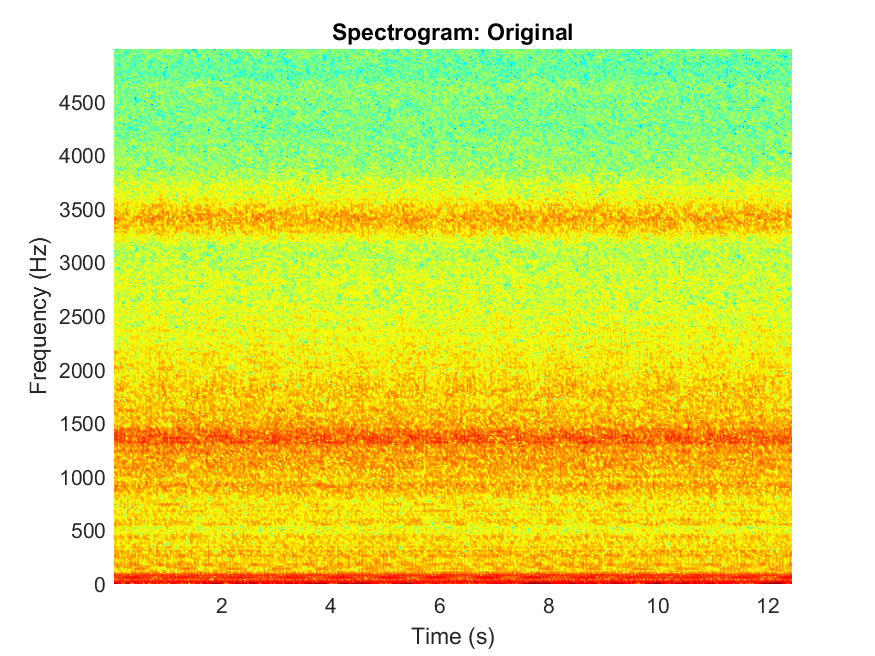}%
\label{spec21_1c}}
\hspace{5mm}
\subfloat[]{\includegraphics[width=1.6in]{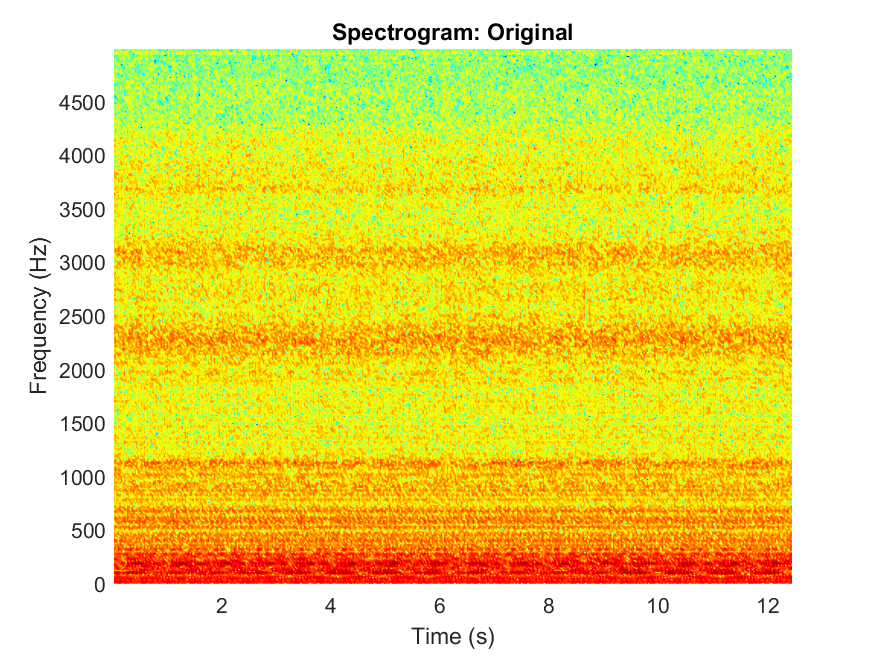}%
\label{spec21_1d}}
\caption{Spectrogram of $S_{21}$ for different fault conditions: (a) Normal  (b) Imbalance (c) Inner race fault (d) Outer race fault}
\label{spec21}
\end{figure}

Three Omni-directional antennas are chosen to operate at 433 MHz, 2.4 GHz, and 5.8 GHz having lengths 11.5 cm, 10.6 cm, and 17.2 cm, respectively. The bearing block closest to the motor is chosen from where the vibrations are to be measured. The measurements are performed at four operating conditions: normal, inner race fault, outer race fault, and imbalance condition. 

For the measurement of the reflection coefficient and transmission coefficient, the placement of the antenna is shown in Fig. \ref{setup1}. $S_{11}$ measurement employs one antenna and uses reflection from the antenna due to impedance mismatch as in Fig. \ref{setup_1a}. $S_{21}$ measurement uses two antennas and records the transmission of the emitted signal from antenna-1 to antenna-2 as in Fig. \ref{setup_1b}. The S-parameters are measured using a vector network analyzer (VNA) under continuous-time mode for the aforementioned operating conditions and the time-domain signal is extracted using Inverse FFT. The $S_{11}$ and $S_{21}$ measurements are performed separately in an anechoic chamber. To study the effect of antenna distance from the source of vibration on the classification performance, S-parameters are measured at 0 cm, 5 cm, and 10 cm from the side of the bearing block. The antenna positions shown in Fig. \ref{setup1} is considered to be at 0 cm. The total number of datasets collected were 2880 (4 operating conditions $\times$ 3 frequencies $\times$ 3 positions $\times$ 40 trials $\times$ 2), where $S_{11}$ and $S_{21}$ comprises of 1440 datasets each. Each of the S parameters is measured for 5 seconds for the operating conditions. The measured reflection coefficient and transmission coefficient are shown in Fig. \ref{raw11} and Fig. \ref{raw21} respectively. Then the spectrograms of the reflection and transmission coefficient are generated using Fast Fourier transform applied to the time-domain signals. Example of spectrograms for measured $S_{11}$ and $S_{21}$ are shown in Fig. \ref{spec11} and Fig. \ref{spec21} respectively. In this study only the magnitude  of the S-parameter is considered as it contains rich time-varying features compared to phase data.

The variation of S11 is by the disruption of the antenna’s near reactive field by the fault-specific vibrations. In the case of S21, it is by the amount of power received at the receiver relative to the input power at Tx. The input power to the Tx is supplied using a two-port vector network analyzer. As the various fault induces a different vibration response, it results in different S11 and S21 response.

The radius of the reactive near field region around the antenna can be calculated using Eqn. \ref{eqn1}. 
\begin{equation}
    Reactive\ near\ field < 0.62\sqrt{\frac{L^{3}}{\lambda }}
    \label{eqn1}
\end{equation}
where L is the largest dimension of the antenna and $\lambda$ is the operating frequency. In the reactive near field region, there is an accumulation of energy near the antenna. As an external conductor passes through this area, field energy is transferred to the conductor's electrons, causing the antenna to lose energy. The antenna impedance shifts as a result of this effect. Thus, the vibration produces a shift in the reflection coefficient over time.

\begin{figure}[!h]
    \centering
    \includegraphics[width=4.5 in]{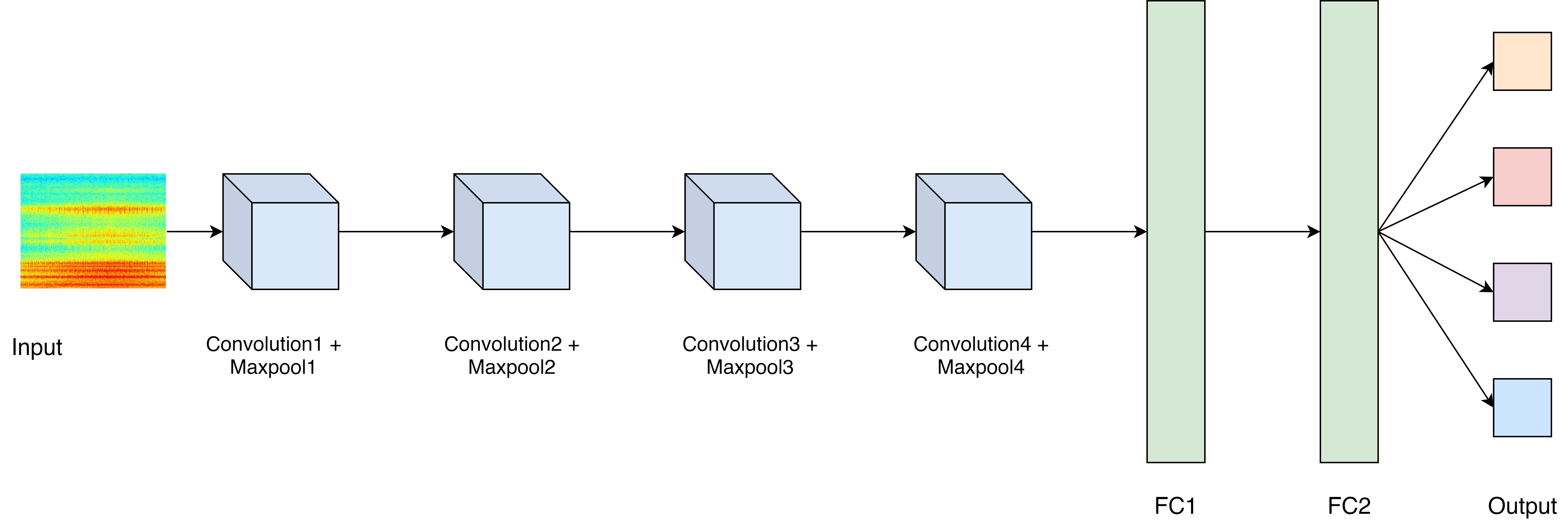}
    \caption{DCNN architecture}
    \label{dcnn}
\end{figure}

\section{Classification}

To classify the different conditions based on the spectrograms of the S parameters, we have employed DCNN due to its superior performance in terms of classification accuracy and computational complexity compared to other techniques such as principal component analysis \citep{c11}, linear predictive coding \citep{c10}, and empirical mode decomposition \citep{c12}. Deep learning exceeds previous machine learning algorithms greatly in image or speech recognition \citep{algo1,algo2,algo3}. DCNN treats the classification of spectrograms as an image recognition problem, and despite the location change, it is capable of identifying signatures. The deep learning approach is ideal because the spectrogram data we aim to identify can be considered as an image. 

The core components of a DCNN are the convolutional layer, pooling layer, and fully connected layer. The feature extraction and dimension reduction are performed by convolution and pooling layer, and the final output is generated using a fully connected layer. In this study, the DCNN consists of 4 layers, where each layer consists of a convolution layer followed by a pooling layer and the activation function used is ReLU. Then for the final classification stage, we use the fully connected layer. The number of convolutional filters used in the first and second layer is 96 and in the third and fourth layer is 256. The size of the convolutional filters is kept fixed at 3$\times$3 in all the layers. A reduction ratio of 2:1 is implemented in the pooling layers. The implemented DCNN structure is shown in Fig. \ref{dcnn}.

\section{Results and Discussion}

For the classification, multiple spectrogram datasets are created for different time durations of the measured vibration signal. We investigate the performance of the DCNN in classifying different fault conditions for different duration of the signal. The aim is to find out the minimum time duration, which would give the highest classification accuracy. From the dataset, 70\% of the data is used for training, and the remaining 30\% is used for validation. The size of the spectrogram is kept fixed at 80$\times$80. Deep network designer toolbox in MATLAB is used for designing and training DCNN.

\begin{figure}[!h]
    \centering
    \includegraphics[width=3 in]{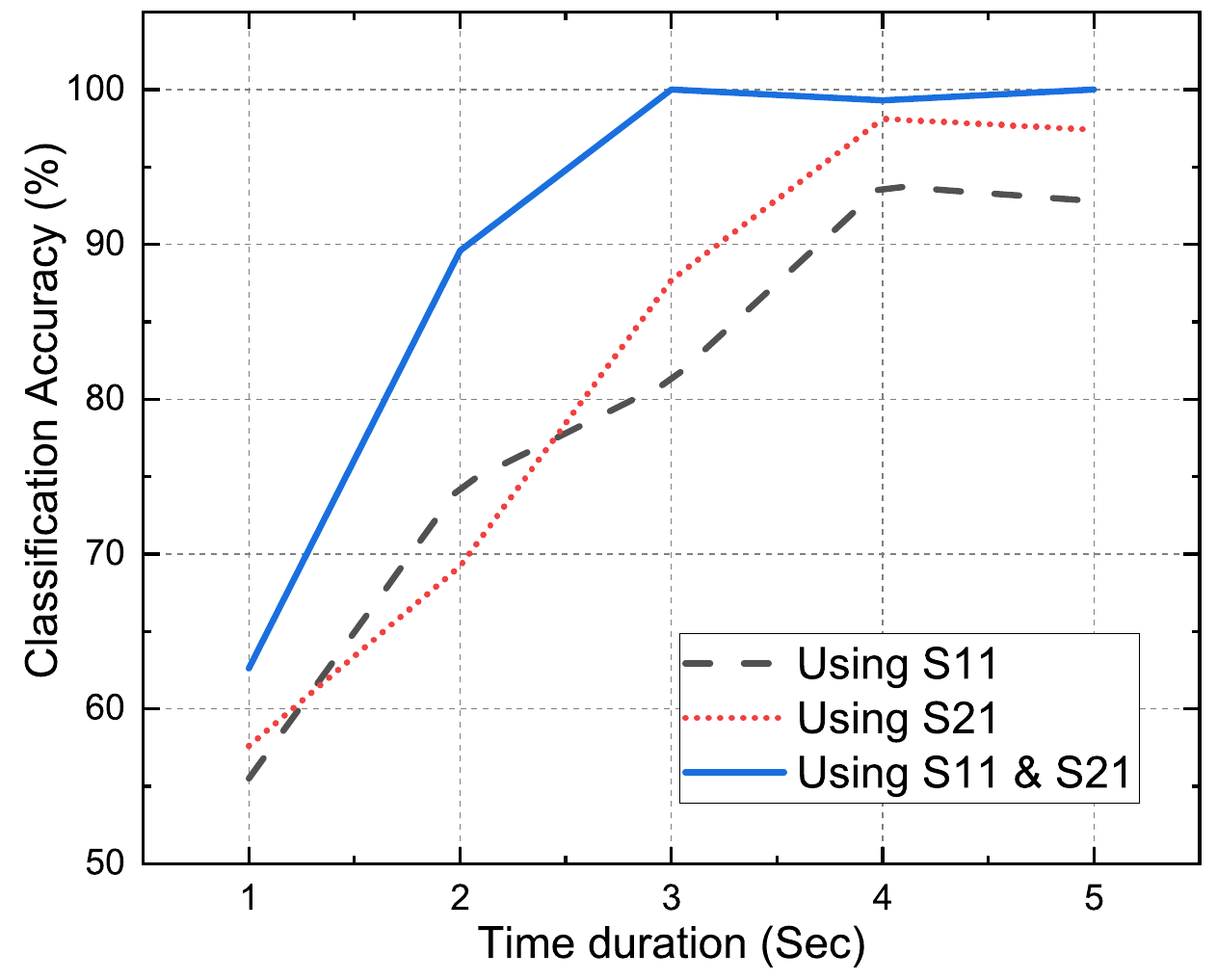}
    \caption{Classification accuracy vs time duration}
    \label{time}
\end{figure}

Fig. \ref{time} shows the classification accuracy vs. time duration of the signal for three cases: a) using $S_{11}$, b) using $S_{21}$ and c) using $S_{11}$ and $S_{21}$ both. For the third case, both the spectrograms are merged together having size 160$\times$80. It is observed from Fig. \ref{time} that classification accuracy of 93\% and 98.1\% is achieved using $S_{11}$ and $S_{21}$ respectively with 4 sec time duration and 100\% accuracy using both $S_{11}$ and $S_{21}$ with 3 sec time duration.

\begin{figure}[!h]
\centering
\subfloat[]{\includegraphics[width=1.6in]{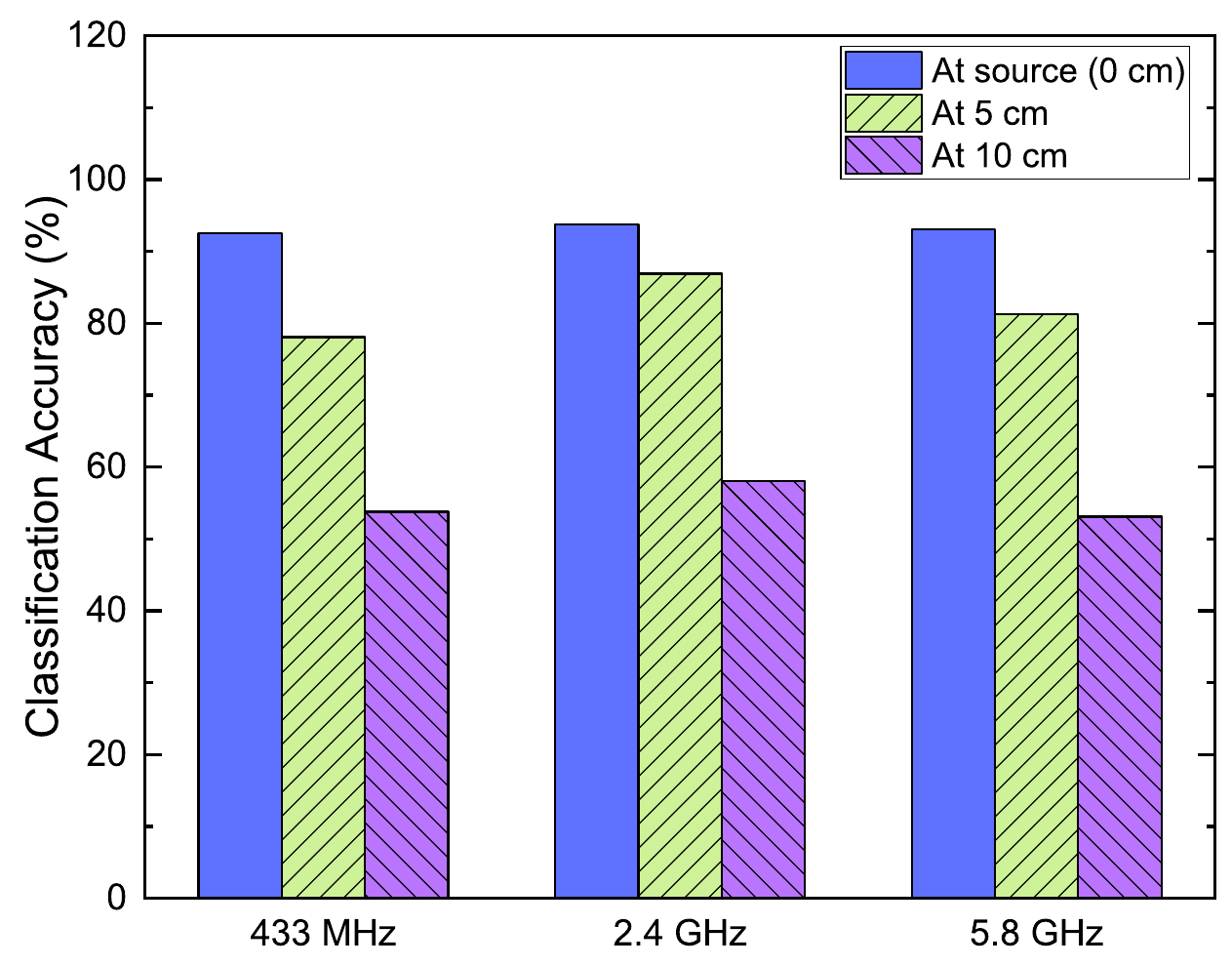}%
\label{r1a}}
\hfill
\subfloat[]{\includegraphics[width=1.6in]{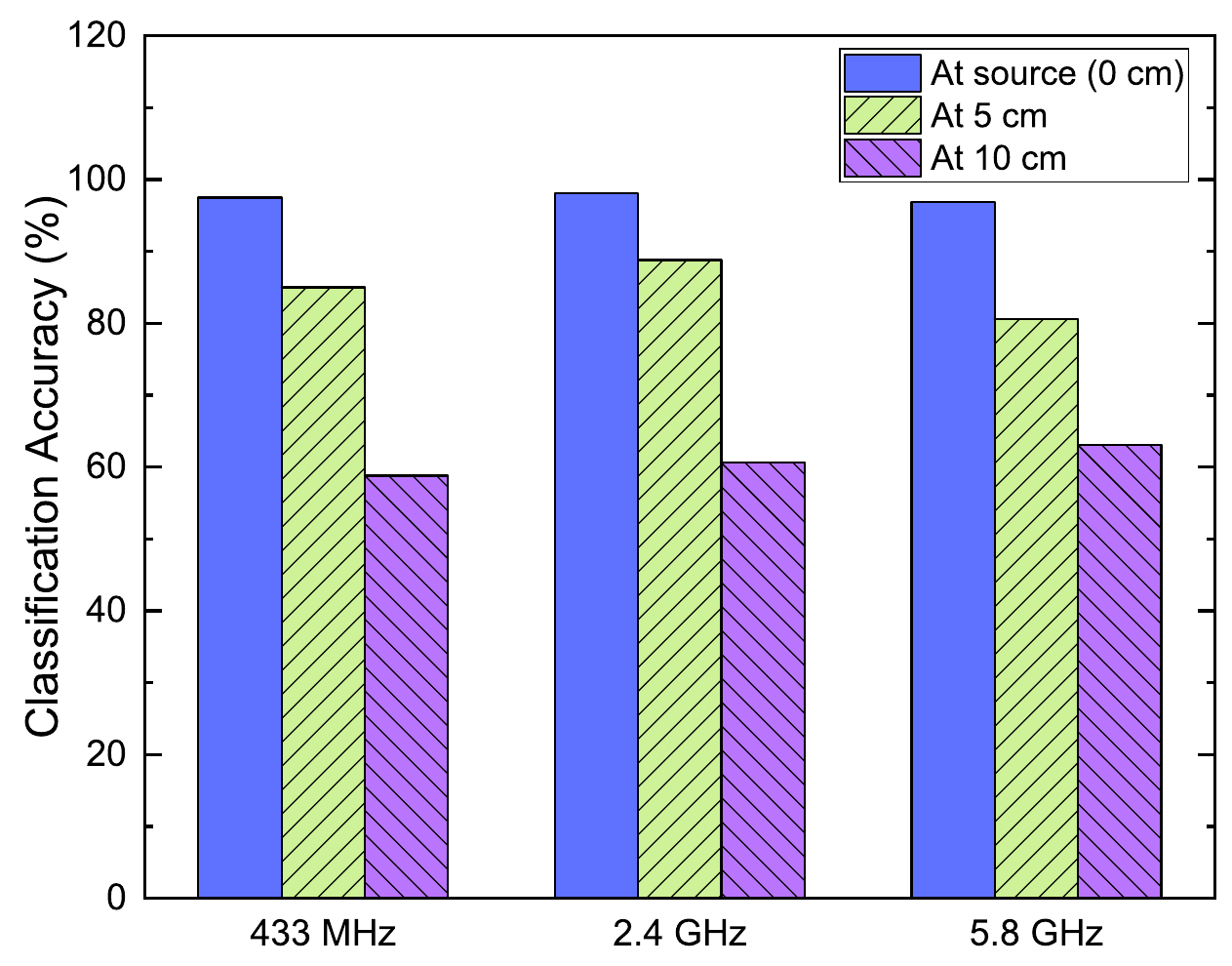}%
\label{r1b}}
\hfill
\subfloat[]{\includegraphics[width=1.6in]{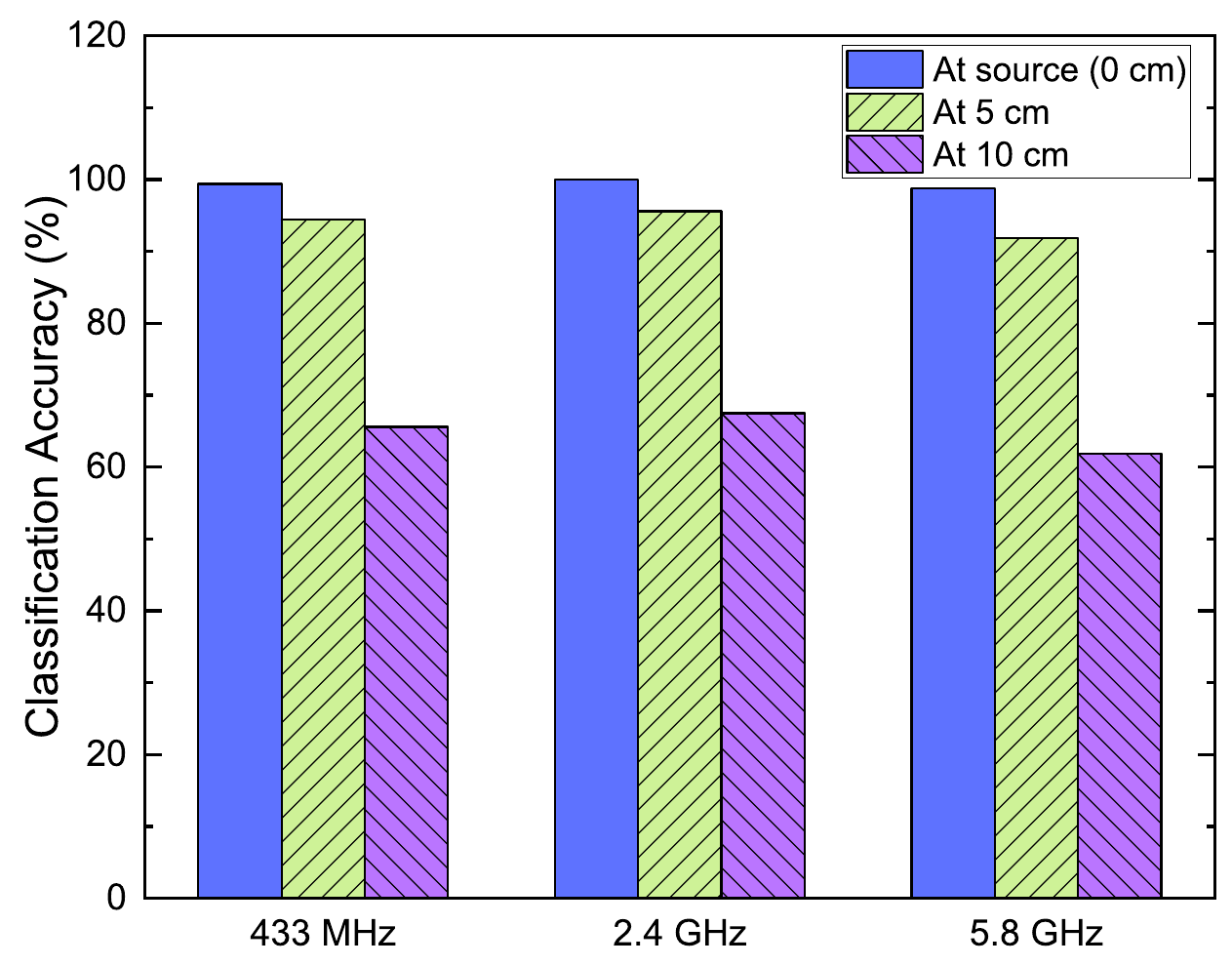}%
\label{r1c}}

\caption{Classification accuracy at different locations at three operating frequency (a) Using $S_{11}$  (b) Using $S_{21}$ (c) Using $S_{11}$ and $S_{21}$}
\label{rr}
\end{figure}

\begin{figure}[!h]
\centering
\subfloat[]{\includegraphics[width=1.6in]{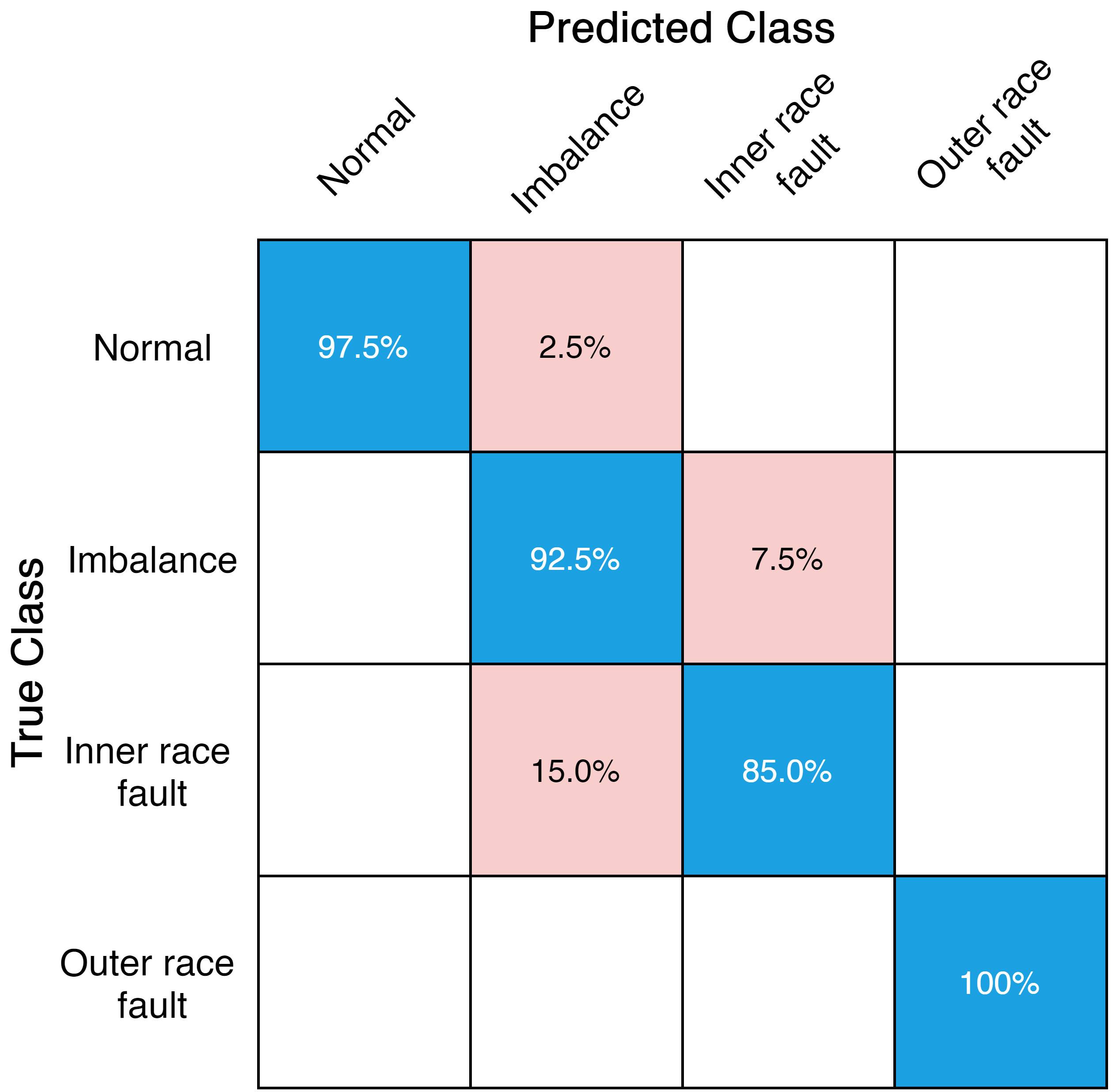}%
\label{confusion1}}
\hfill
\subfloat[]{\includegraphics[width=1.6in]{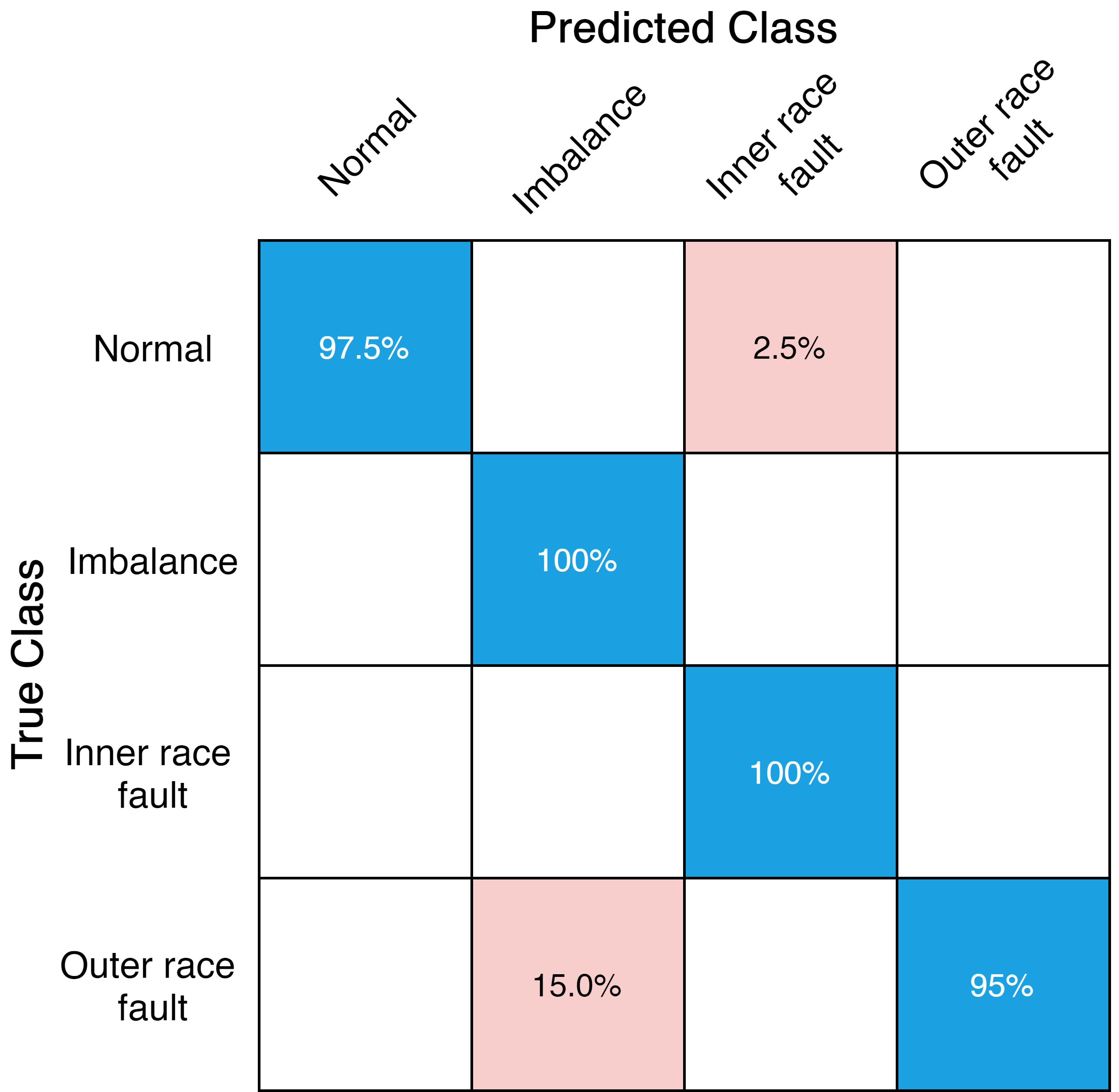}%
\label{confusion2}}
\hfill
\subfloat[]{\includegraphics[width=1.6in]{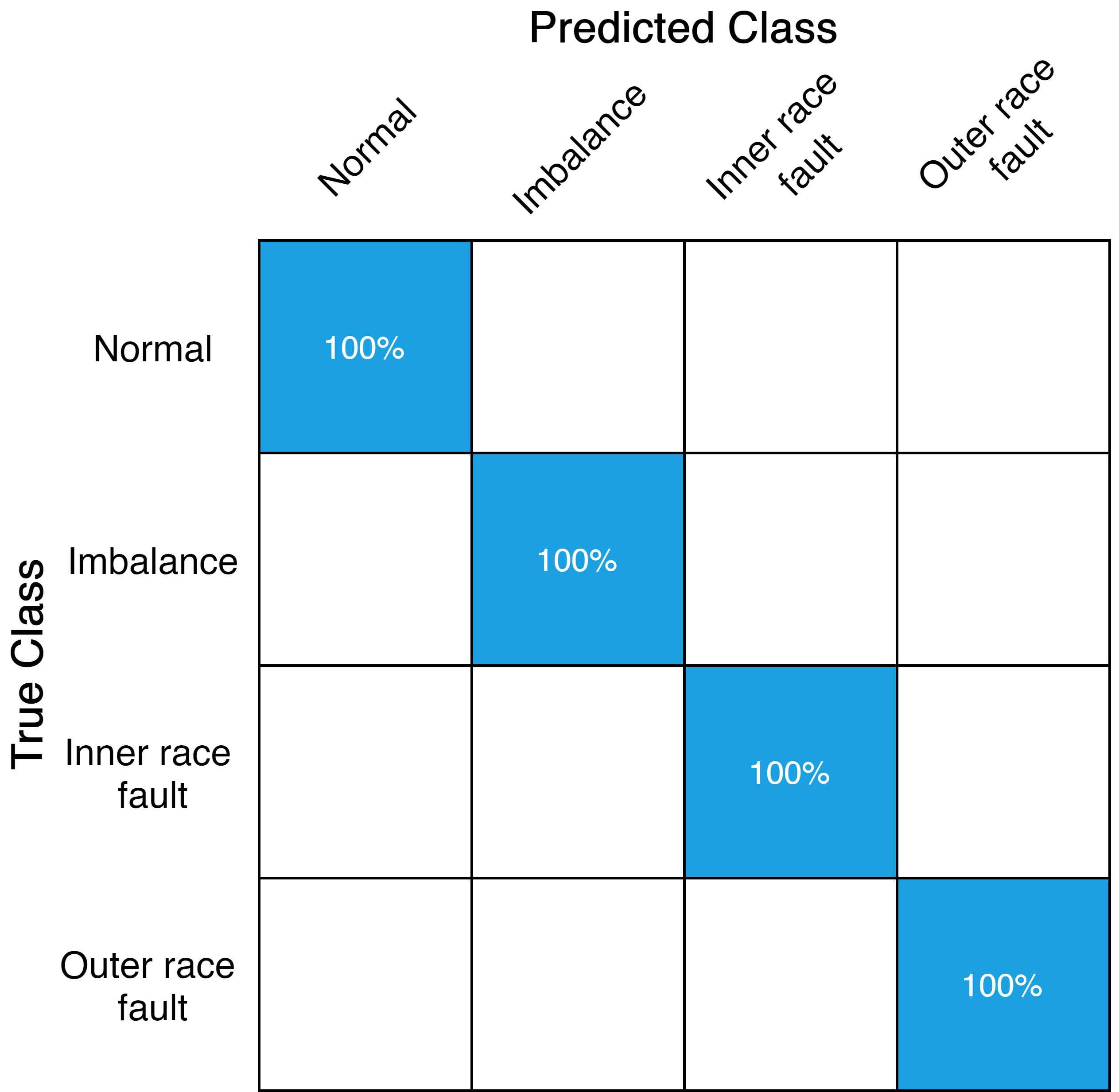}%
\label{confusion3}}

\caption{Confusion matrix for 2.4 GHz antenna placed at source (a) Using $S_{11}$  (b) Using $S_{21}$ (c) Using $S_{11}$ and $S_{21}$}
\label{confusion}
\end{figure}

Furthermore, the effect of antenna distance from the source on the classification performance is also investigated. Figure \ref{rr} shows the classification accuracy using three different operating frequencies at three different locations. It is observed that the classification accuracy decreases as the antenna moves away from the source. When the antenna is placed at the source, the classification performance is maximum. However, while measuring the $S_{11}$, it is important to consider the radius of the antenna's near-reactive field so that the vibrating source is within the reactive field of the antenna. The radius of the near-reactive field for the three antennas is 2 cm for 433 MHz, 6 cm for 2.4 GHz and 19 cm for 5.8 GHz. In this study, 2.4 GHz has an optimal radius for the near-reactive field.
\begin{table}[!ht]
\centering
\caption{Performance metrics for the three case }
\begin{tabular}{@{}lcccc@{}}
\toprule
            & \multicolumn{1}{l}{Accuracy} & \multicolumn{1}{l}{Precision} & \multicolumn{1}{l}{Recall} & \multicolumn{1}{l}{F1 Score} \\ \midrule
S11         & 93.8\%                       & 94\%                          & 93.75\%                    & 93.75\%                      \\
S21         & 98.1\%                       & 98.25\%                       & 98\%                       & 98.25\%                      \\
S11 and S21 & 100\%                        & 100\%                         & 100\%                      & 100\%                        \\ \bottomrule
\end{tabular}
\label{metric}
\end{table}
Figure \ref{confusion} shows the confusion matrix for the three cases when the 2.4 GHz antenna is at the source. It is observed that the highest confused conditions are inner race fault and imbalance as in Fig. \ref{confusion1} and outer race fault and imbalance as in Fig. \ref{confusion2}. Whereas the confusion is not there when both the $S_{11}$ and $S_{21}$ is used as shown in Fig. \ref{confusion3}. Table \ref{metric} shows the performance metric of the DCNN.

Zhang et. al \citep{zhang2020} reported comprehensive analysis for diagnosis of induction motor fault using sensors like accelerometer, current sensor and thermocouple, based on deep learning algorithms where the average accuracy of classification is 95\%. Comparably, our method has obtained a high accuracy of 100\%, which means that the classification based on our approach is effective.

\section{Conclusion}

In this paper, we have applied DCNN for the classification of different motor faults based on the spectrogram of $S_{11}$ and $S_{21}$ of antenna. We collected the data using Omni-directional antennas operating at frequencies 433 MHz, 2.4 GHz and 5.8 GHz at three different locations. It is observed that the classification accuracy decreases as the antenna moves away from the source of vibration for both the case of $S_{11}$ and $S_{21}$. A classification accuracy of 93\% and 98.1\% is achieved using $S_{11}$ and $S_{21}$ respectively and 100\% accuracy using both $S_{11}$ and $S_{21}$. The effect of time duration of the signal is also investigated and it is found that the highest accuracy is achieved with a minimum time duration of 3 seconds. In addition, compared to the previous deep learning-based approach \citep{zhang2020}, we believe that our results show potential for motor fault classification based on the antenna approach using DCNN.

Future works can include different directional antennas, other motor-related faults, diverse antenna placements at different working environments, and implementation in real-time. It is also necessary to highlight the method's limitations. Because the shape of the signature in a spectrogram is critical to classification, variations caused by irregular vibrations can degrade performance. If the antenna is placed too far away from the vibration source, its performance will suffer as well. Furthermore, the computational complexity of DCNN is typically higher than that of data-driven models from traditional machine learning algorithms due to its multilayer structure. As a result, the computational complexity of DCNN should be carefully considered in applications that require real-time processing.

\section*{Declaration of Competing Interest}
The authors declare that they have no known competing financial interests or personal relationships that could have appeared to influence the work reported in this paper.

\bibliographystyle{model5-names}
\biboptions{authoryear}
\bibliography{reference}

\end{document}